\documentclass[%
 reprint,
 superscriptaddress,
 amsmath,amssymb,
 aps,
 amsfonts,
 prl,
]{revtex4-2}

\usepackage{graphicx}
\usepackage{dcolumn}
\usepackage{bm}
\usepackage{xcolor}
\usepackage{braket}
\usepackage{mathrsfs}
\usepackage[version=4]{mhchem} 
\usepackage{hyperref} 
\hypersetup{
    colorlinks=true,
    linkcolor=red, 
    urlcolor=blue,  
    citecolor=blue, 
    }
\newcommand{\DefineAuthor}[2]{%
  \expandafter\newcommand\csname #1note\endcsname[1]{%
    \textbf{\textcolor{#2}{\textbf{#1:} ##1}}}%
  \expandafter\newcommand\csname #1\endcsname[1]{
    \textbf{\textcolor{#2}{##1}}}
  \expandafter\newcommand\csname #1cancel\endcsname[1]{%
    \textbf{\textcolor{#2}{\sout{##1}}}}%
  \expandafter\newcommand\csname #1change\endcsname[2]{%
    \textbf{\textcolor{#2}{\sout{##1} ##2}}}%
  \newenvironment{#1text}{\color{#2}}{\color{black}}
}
\definecolor{dartmouthgreen}{rgb}{0.05, 0.5, 0.06}
\DefineAuthor{ER}{magenta}
\DefineAuthor{GK}{green}
\DefineAuthor{HK}{purple}
\DefineAuthor{ZJA}{orange}
\DefineAuthor{RAD}{blue}
\DefineAuthor{NAB}{red}
\DefineAuthor{GEN}{dartmouthgreen}
\DefineAuthor{FIG}{red}
\usepackage{soul}

\newcommand{\ie}[0]{\textit{i.e.}, }
\newcommand{\eg}[0]{\textit{e.g.}, }
\newcommand{\via}[0]{\textit{via} }
\newcommand{\cf}[0]{\textit{cf}. }
\makeatletter 
\providecommand*{\diff}%
        {\@ifnextchar^{\DIfF}{\DIfF^{}}}
\def\DIfF^#1{%
        \mathop{\mathrm{\mathstrut d}}%
            \nolimits^{#1}\gobblespace}
\def\gobblespace{%
        \futurelet\diffarg\opspace}
\def\opspace{%
        \let\DiffSpace\!%
        \ifx\diffarg(%
            \let\DiffSpace\relax
        \else
            \ifx\diffarg[%
                \let\DiffSpace\relax
            \else
                \ifx\diffarg\{%
                    \let\DiffSpace\relax
                \fi\fi\fi\DiffSpace}
\begin{document}

\title{Orbitally-Resolved Mechanical Properties of Solids from Maximally Localized Wannier Functions}

\author{Ethan T. Ritz}
\thanks{These authors contributed equally to this work.}
\affiliation{Sibley School of Mechanical and Aerospace Engineering, Cornell University, Ithaca, NY 14853, USA}
\author{Guru Khalsa}
\thanks{These authors contributed equally to this work.}
\affiliation{Department of Materials Science and Engineering, Cornell University, Ithaca, NY 14853, USA}
\author{Hsin-Yu Ko}
\affiliation{Department of Chemistry and Chemical Biology, Cornell University, Ithaca, NY 14853, USA}
\author{Ju-an Zhang}
\affiliation{Department of Chemistry and Chemical Biology, Cornell University, Ithaca, NY 14853, USA}
\author{Robert A. DiStasio Jr.}
\email{distasio@cornell.edu}
\affiliation{Department of Chemistry and Chemical Biology, Cornell University, Ithaca, NY 14853, USA}
\author{Nicole A. Benedek}
\email{nbenedek@cornell.edu}
\affiliation{Department of Materials Science and Engineering, Cornell University, Ithaca, NY 14853, USA}


\begin{abstract} 
\noindent We present a technique for partitioning the total energy from a semi-local density functional theory calculation into contributions from individual electronic states in a localized Wannier basis. We use our technique to reveal the key role played by the $s$ and $p$ orbitals of the apical oxygen atoms in a curious elastic anomaly exhibited by ferroelectric \ce{PbTiO3} under applied stress, which has so far gone unexplained. Our technique enables new insights into the chemical origins of the mechanical properties of materials, or any property given by an energy derivative. 
\end{abstract}

\maketitle

\textit{Introduction}.\ $\mathrm{-}$ Almost all of the functional properties of materials depend on, or are strongly affected by, their mechanical properties. For example, the significant lattice stresses induced during repeated charging/discharging cycles lead to rapid capacity fade in Li-ion battery materials~\cite{manthiram20}. The interplay between the elastic constants and vibrational properties of a material determines the rate at which it expands or contracts upon heating or cooling~\cite{burg16}. The small molecule adsorption capacity of metal-organic frameworks is maximized for materials that are mechanically highly flexible~\cite{mason15}. Theories describing the mechanical properties of materials (\eg linear/non-linear elasticity, plasticity, fracture) have been developed, and it is now possible in many cases to calculate the values of mechanical constants using routine computational techniques~\cite{dejong15}. However, knowing the value of a particular elastic constant (say), does not necessarily tell us anything about how that value arises from microscopic chemical and physical factors. This lack of knowledge hinders the design and synthesis of novel materials with tailored mechanical properties.

There is a long history of efforts to link the elastic properties of materials to chemically intuitive concepts related to structure and bonding. For example, one of Pauling's rules for ionic crystals is that shorter bonds are generally stronger bonds, although it has also been applied to other materials families~\cite{gibbs98,grochala07,zeier16}. Empirical scaling relations have been derived for the bulk moduli of metals, semiconductors, and ionic crystals~\cite{cohen85}, an approach that usually involves fitting parameters to large sets of experimental (and sometimes theoretical) data on known materials. Tight-binding methods have been used extensively to understand the contributions of specific bonds to the elastic properties of semiconductors~\cite{harrison}. The key strengths of these approaches are that they coarse grain away the details such that ``big picture'' property trends can be synthesized across many different materials. However, it is often difficult to know in advance which details really matter. Hence, a scheme that combines intuitive and local notions of structure and bonding with the microscopic insights available from first-principles calculations would be highly desirable.

In this Letter, we take the first steps towards developing a first-principles approach that links macroscopic mechanical properties to local, chemical-specific materials features, such as composition, structure, and bonding. Our approach starts with transforming the Bloch states for a material of interest (computed using semi-local density functional theory (DFT)) into a basis of maximally localized Wannier functions (MLWFs)~\cite{marzari97,marzari12}. Since the MLWF basis is orthonormal and complete, derivatives of the total energy (\emph{all} of the energy, not just the band energy, as in previous approaches~\cite{majewski1986crystal}) are then partitioned into a series of \textit{orbitally-resolved} electronic contributions and \textit{ionically-resolved} ion-ion contributions. This key and novel step of our approach maps the computed values of mechanical constants (or any property given by an energy derivative) onto individual orbital and ion contributions, and can be used to gain new insights into the chemical origins of the mechanical properties of complex materials.

We demonstrate the power and utility of our approach by using it to gain a new understanding of the anomalous elastic behavior of ferroelectric \ce{PbTiO3}, a material of both fundamental scientific and technological importance. \ce{PbTiO3} is an exceptional piezoelectric and one of the few known materials to undergo \emph{volumetric} negative thermal expansion over an appreciable temperature range~\cite{shirane1951phase,chen2005structure,ritz2018}. A number of prior studies have shown that \ce{PbTiO3} exhibits a curious elastic anomaly whereby the $c$-axis lattice parameter in the ferroelectric tetragonal phase suddenly jumps under negative hydrostatic pressure or small positive applied stress~\cite{tinte03,duan08,moriwake08,sharma14,setter16}. This jump coincides with `kinks' in the values of the elastic compliance constants and piezoelectric strain coefficients. Despite concerted study over the past two decades, the microscopic origin of this behavior has not been identified. We use the technique described here to reveal the key role played by the $s$ and $p$ orbitals of the apical oxygen atoms, which in contrast with other states, experience a significant change in internal stress as the applied stress increases. Going further, we correlate changes in the apical oxygen bonding environment with the position of the anomaly in a range of perovskites. This work illustrates how macroscopic mechanical properties can be linked with microscopic, orbital-level features, a necessary step towards enabling atomic-scale design of materials that exhibit novel mechanical behaviors. Finally, we demonstrate that our technique can also be used to gain insight into other properties by considering the ferroelectric phase transition of \ce{BaTiO3}. We expect our technique to become an essential tool for rigorously exploring and understanding materials properties given by energy derivatives. 

\textit{Theory/Method}.\ $\mathrm{-}$ Strategies for partitioning the total or band energy of a system onto a local representation have been previously attempted~\cite{rahm2016distinguishing,baba2006natural,maintz2016lobster,oliphant25}, and the concept of the total energy expressed in whole (or in part) as an integral over the energy density~\cite{cohen2000total,nakai2007extension} or sum over electronic bands~\cite{cohen1994tight} or elements of the Hamiltonian~\cite{finnis2007bond} has been explored as an analytical tool. The method proposed here is distinguished from prior efforts in several respects: it partitions the total energy, it is suitable for analyzing multi-element crystalline materials, and it employs a basis that is orthonormal, complete, and amenable to chemical interpretation.

We start with the total Kohn--Sham (KS) DFT energy, $E_{\rm tot} = E_{\rm kin} + E_{\rm H} + E_{\rm xc} + E_{\rm ext}$, in which $E_{\rm kin}$, $E_{\rm H}$, and $E_{\rm xc}$ are the KS kinetic, Hartree (H), and exchange-correlation (xc) energies, and $E_{\rm ext} = E_{\rm ext}^{\rm (e-I)} + E_{\rm ext}^{\rm (I-I)} + E_{\rm ext}^{\rm (e/I-F)}$ is the external energy, which includes electron-ion (e$-$I) and ion-ion (I$-$I) interactions, as well as electron-/ion- interactions with any applied fields (e/I$-$F, omitted here for simplicity). Throughout this work, all energies are computed using semi-local DFT (an extension to hybrid DFT will be considered in future work), ions refer to nuclei and their associated frozen-core electrons (modeled using the pseudopotential approach~\cite{Pickett1989,psp_review2011}), and all references to $E_{\rm H}$, $E_{\rm ext}^{\rm (e-I)}$, and $E_{\rm ext}^{\rm (I-I)}$ correspond to finite-valued quantities with individual divergences (at infinite system sizes) removed using standard regularization techniques~\cite{supp,de_Leeuw1980,Makov1995}. 

We seek to partition $E_{\rm tot}$ into an \textit{orbitally resolved} electronic energy ($E_{\rm elec} \equiv E_{\rm tot} - E_{\rm ext}^{\rm (I-I)}$) and an \textit{ionically resolved} ion-ion interaction energy ($E_{\rm ext}^{\rm (I-I)}$). To proceed, we first note that the orbitals in a strict KS scheme are subject to the same potential in the electronic Hamiltonian, $\hat{H}_{\rm elec} = -\frac{1}{2}\nabla^2 + v_{\rm H}(\mathbf{r}) + v_{\rm xc}(\mathbf{r}) + v_{\rm ext}(\mathbf{r})$, in which $v_{\rm H}(\mathbf{r}) = \delta {E}_{\rm H}/\delta n(\mathbf{r})$, $v_{\rm xc}(\mathbf{r}) = \delta E_{\rm xc}/\delta n(\mathbf{r})$, and $v_{\rm ext}(\mathbf{r}) = \delta {E}_{\rm ext}/\delta n(\mathbf{r}) = \delta {E}_{\rm ext}^{\rm (e-I)}/\delta n(\mathbf{r})$ are the Hartree, xc, and electron-ion interaction potentials.
During a standard KS-DFT calculation, the $\hat{H}_{\rm elec}$ eigensystem is solved for the canonical Bloch states $\{ \psi_{\nu\mathbf{k}}(\mathbf{r}) \}$ and energies $\{ \epsilon_{\nu\mathbf{k}} \}$, in which $\nu$ is the band index for a wave vector $\mathbf{k}$ in the first Brillouin zone (BZ, sampled with $N_{k}$ $\mathbf{k}$-points).
In the Bloch representation, $E_{\rm elec}$ can be written as: 
\begin{align}
    E_{\rm elec} &= \frac{1}{N_{k}} \sum_{\nu\mathbf{k}} \epsilon_{\nu\mathbf{k}} - \int_{\Omega} \diff\mathbf{r} \, n(\mathbf{r}) \Delta v(\mathbf{r}) ,
    \label{eqn:total_energy_KS_double_counting}
\end{align}
in which $\Omega$ is the unit cell, $n(\mathbf{r})$ is the electron density, $\Delta v(\mathbf{r}) \equiv \frac{1}{2} v_{\rm H}(\mathbf{r}) + v_{\rm xc}(\mathbf{r}) - \epsilon_{\rm xc}(\mathbf{r})$, and $\epsilon_{\rm xc}(\mathbf{r})$ is the xc energy density.

While $E_{\rm elec}$ in Eq.~\eqref{eqn:total_energy_KS_double_counting} can be \textit{band resolved} by partitioning the sum over $\nu$ in both $\sum_{\nu\mathbf{k}} \epsilon_{\nu\mathbf{k}}$ and $n(\mathbf{r}) = \frac{1}{N_{k}} \sum_{\nu\mathbf{k}} \left| \psi_{\nu\mathbf{k}}(\mathbf{r}) \right|^2$, and then averaging over $\mathbf{k}$ points, \ie $E_{\mathrm{elec}} = \sum_{\nu} \langle E_{\nu} \rangle \equiv \sum_{\nu} \frac{1}{N_{k}} \sum_{\mathbf{k}} E_{\nu\mathbf{k}}$ with $E_{\nu\mathbf{k}} \equiv \epsilon_{\nu\mathbf{k}} - \int_{\Omega} \diff\mathbf{r} \, |\psi_{\nu\mathbf{k}}(\mathbf{r})|^2 \Delta v(\mathbf{r})$, the Bloch states are typically delocalized and therefore difficult to assign to a particular atom, bond, or functional group.
Since our aim is to construct a rigorous orbitally resolved $E_{\rm elec}$ that enables one to extract chemical insight into the mechanical properties of materials, we adopt a localized orbital representation ($\{ w_{\alpha\mathbf{R}}(\mathbf{r}) \}$; in this work, we use maximally localized Wannier functions (MLWFs)~\cite{marzari97,marzari12}) obtained \via a unitary rotation of $\{ \psi_{\nu\mathbf{k}}(\mathbf{r}) \}$, \ie $w_{\alpha\mathbf{R}}(\mathbf{r}) = \frac{1}{N_{k}}\sum_{\mathbf k}e^{-i \mathbf{k} \boldsymbol{\cdot} \mathbf{R}} \sum_{\nu} U^{(\mathbf{k})}_{\alpha \nu} \psi_{\nu\mathbf{k}}(\mathbf{r})$. In this expression, $\mathbf{R}$ is one of the $N_k$ Bravais lattice vectors in the Born--von Karman supercell ($\mathbb{S}$, which is $\Omega$ replicated with respect to the $\mathbf{k}$-point mesh) and $U^{(\mathbf{k})}_{\alpha \nu}$ is a unitary matrix element for a given $\mathbf{k}$. Since $E_{\rm tot}$ is invariant to such transformations~\cite{marzari97,marzari12} and $E_{\rm ext}^{\rm (I-I)}$ is orbital-independent, $E_{\rm elec} = E_{\rm tot} - E_{\rm ext}^{\rm (I-I)}$ is also rigorously conserved when expressed in the localized MLWF representation, \ie $E_{\rm elec} = \frac{1}{N_{k}} \left[ \sum_{\alpha\mathbf{R}} \epsilon_{\alpha\mathbf{R}} - \int_{\mathbb{S}} \diff\mathbf{r} \, n(\mathbf{r}) \Delta v(\mathbf{r})\right]$ with $\epsilon_{\alpha\mathbf{R}} \equiv \braket{w_{\alpha\mathbf{R}}(\mathbf{r})| \hat{H}_{\rm elec} |w_{\alpha\mathbf{R}}(\mathbf{r})}$ and $n(\mathbf{r}) = \sum_{\alpha\mathbf{R}}\left| w_{\alpha\mathbf{R}}(\mathbf{r}) \right|^2$. 

In the localized MLWF representation, we can define the following quantity (in analogy to $E_{\nu\mathbf{k}}$ above), 
\begin{align}
    E_{\alpha \mathbf{R}} \equiv \epsilon_{\alpha\mathbf{R}} - \int_{\mathbb{S}} \diff\mathbf{r} \, \left| w_{\alpha\mathbf{R}}(\mathbf{r}) \right|^2 \Delta v(\mathbf{r}) ,
\end{align}
from which an \textit{orbitally resolved} $E_{\rm elec}$ can be written as:
\begin{align}
    E_{\rm elec} &= \frac{1}{N_{k}} \sum_{\alpha\mathbf{R}} E_{\alpha \mathbf{R}} = \sum_{\alpha} E_{\alpha \mathbf{0}} .
    \label{eqn:electronic_energy_wannier}
\end{align}
Since $w_{\alpha\mathbf{R}}(\mathbf{r}) = w_{\alpha\mathbf{0}}(\mathbf{r}-\mathbf{R})$ and both $\hat{H}_{\rm elec}$ and $\Delta v(\mathbf{r})$ are translationally invariant (with respect to $\mathbf{R}$), $\epsilon_{\alpha\mathbf{R}}=\epsilon_{\alpha\mathbf{0}}$, $\int_{\mathbb{S}} \diff\mathbf{r} \, \left| w_{\alpha\mathbf{R}}(\mathbf{r}) \right|^2 \Delta v(\mathbf{r}) = \int_{\mathbb{S}} \diff\mathbf{r} \, \left| w_{\alpha\mathbf{0}}(\mathbf{r}) \right|^2 \Delta v(\mathbf{r})$, and hence $E_{\alpha \mathbf{R}} = E_{\alpha \mathbf{0}}$.
The second equality in Eq.~\eqref{eqn:electronic_energy_wannier} then directly follows as $\sum_{\mathbf{R}} = N_k$.
This MLWF-based partitioning of $E_{\rm elec}$ is particularly amenable to chemical analysis in complex materials as it is \textit{fully} orbitally resolved (and does not involve $\mathbf{k}$-point averaged quantities) and based on localized orbitals that can be assigned to a particular atom, bond, or functional group.
This partitioning is also practically convenient as one only needs to consider the single branch of MLWFs ($\{w_{\alpha\mathbf{0}}(\mathbf{r})\}$) centered in $\Omega$.
%
%
\begin{figure*}[t!]
    \centering
    \includegraphics[width=0.99\textwidth]{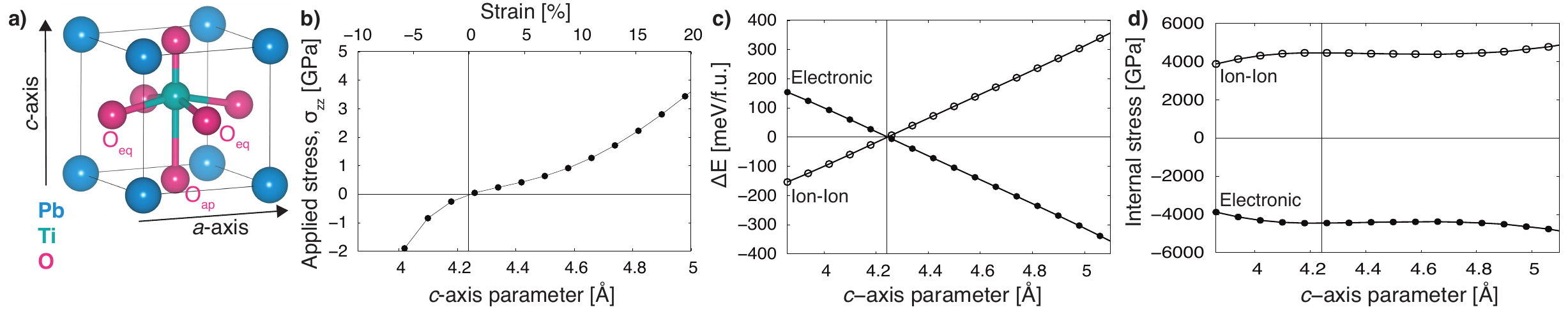}
    \caption{a) Ferroelectric \ce{PbTiO3} in the tetragonal $P4mm$ space group with equatorial (\ce{O_{eq}}) and apical (\ce{O_{ap}}) oxygens highlighted. There is one short and one long \ce{Ti-O_{ap}} bond along the tetragonal $c$-axis. 
    b) Variation in the $c$-axis lattice parameter with applied stress along the tetragonal axis ($\sigma_{zz}$) shows significant non-linearity at small stresses. The strain induced by $\sigma_{zz}$ is plotted on the top $x$-axis. 
    c) Electronic ($E_{\rm elec}$, closed circles) and ion-ion ($E_{\rm ext}^{\rm (I-I)}$, open circles) energy differences between equilibrium \ce{PbTiO3} and \ce{PbTiO3} under applied stress. 
    d) Internal stresses (\cf Eq.~\eqref{eq:internal}) associated with $E_{\rm elec}$ and $E_{\rm ext}^{\rm (I-I)}$ for \ce{PbTiO3} in response to an applied stress. 
    Vertical lines at $c = 4.242$~\AA~denote the computed equilibrium $c$-axis lattice parameter for \ce{PbTiO3}.
    In c) and d), the $c$-axis lattice parameters resulting from the applied stress are plotted on the $x$-axis, rather than the applied stress itself.
    }
    \label{energies}
\end{figure*}
%
%

Having resolved $E_{\rm elec}$ into MLWF-specific contributions in Eq.~\eqref{eqn:electronic_energy_wannier}, our last task is to derive an \textit{ionically resolved} ${E}_{\rm ext}^{\rm (I-I)} = \sum_{A} E_{\mathrm{ext},A}^{\rm (I-I)}$ by partitioning this term into contributions from each ion $A$ in the system~\cite{supp}. 
With an orbitally resolved $E_{\rm elec}$ and an ionically resolved $E_{\rm ext}^{\rm (I-I)}$ in hand, $E_{\rm tot}$ can now be written as:
\begin{align}
    E_{\rm tot} = E_{\rm elec} + E_{\rm ext}^{\rm (I-I)} = \sum_{\alpha} E_{\alpha\mathbf{0}} + \sum_{A} E_{\mathrm{ext},A}^{\rm (I-I)} .
    \label{eq:Etot-alpha-A}
\end{align}
Since many mechanical properties $P$ can be written as derivatives of $E_{\rm tot}$ with respect to some set of deformations $\{q_{i}\}$ (\eg a strain, a phonon mode), one can also decompose $P$ into a sum of orbitally- and ionically-resolved contributions, \ie
\begin{align}
    P = \left( \frac{\partial}{\partial q_{1}} \right) \cdots \left( \frac{\partial}{\partial q_{N}} \right) E_{\rm tot} = \sum_{\alpha} P_{\alpha} + \sum_{A} P_{A} ,
    \label{eq:decomposition}
\end{align}
in which $P_{\alpha} \equiv \left( \frac{\partial}{\partial q_{1}} \right) \cdots \left( \frac{\partial}{\partial q_{N}} \right) E_{\alpha \mathbf{0}}$ is the $\alpha$-th orbital contribution and $P_{A} \equiv \left( \frac{\partial}{\partial q_{1}} \right) \cdots \left( \frac{\partial}{\partial q_{N}} \right) E_{\rm ext,A}^{\rm (I-I)}$ is the $A$-th ion contribution to $P$.
By resolving computed property values into individual orbital and ion contributions, this partitioning technique can be used to directly probe the influence of local, chemical-specific materials features (\eg composition, structure, and bonding) on the mechanical properties of complex materials.

\textit{Computational Details}.\ $\mathrm{-}$ Our method was integrated into \texttt{Quantum ESPRESSO} (\texttt{v6.4.1})~\cite{giannozzi2009quantum,giannozzi2017advanced} and \texttt{Wannier90} (\texttt{v3.0.0})~\cite{pizzi2020wannier90}. All DFT calculations were performed with the PBEsol~\cite{perdew2008restoring} functional, using an $8 \times 8\times 8~\Gamma$-centered $\mathbf{k}$-point mesh, a plane wave kinetic energy cutoff of $100$~Ry, and norm-conserving pseudopotentials from the PseudoDojo project~\cite{van2018pseudodojo}. Since our work involves computing numerical derivatives, all DFT energies were tightly converged to within $1.0 \times 10^{-10}$~Ry during self-consistent field (SCF) calculations. Lattice parameters and internal structural degrees of freedom were converged such that the energy change between steps was less than $1.0 \times 10^{-5}$~Ry and all force components were less than $3.0 \times 10^{-5}$~Ry/Bohr. All cell degrees of freedom were converged to within $5.0 \times 10^{-3}$~GPa.

\textit{Results}.\ $\mathrm{-}$ \ce{PbTiO3} is cubic ($Pm\bar{3}m$) at high temperatures and undergoes a structural phase transition at $760$~K to a tetragonal ($P4mm$) ferroelectric phase~\cite{shirane1951phase,chen2005structure}, which is the focus of this work (Fig.~\ref{energies}a). Fig.~S1 shows that $E_{\rm tot}$ in the immediate vicinity of the equilibrium $c$-axis lattice parameter is quite flat~\cite{supp}; this region corresponds to the sudden jump in the $c$-axis lattice parameter with applied stress ($\sigma_{zz}$) shown in Fig.~\ref{energies}b and noted in earlier studies~\cite{tinte03,duan08,moriwake08,sharma14,setter16}. To gain microscopic insight into this anomalous behavior, we first partition $E_{\rm tot}$ into $E_{\rm elec}$ and $E_{\rm ext}^{\rm (I-I)}$. Fig.~\ref{energies}c shows that $E_{\rm elec}$ ($E_{\rm ext}^{\rm (I-I)}$) decreases (increases) as the $c$-axis increases. 
%
%
\begin{figure}
\centering
\includegraphics[width=0.35\textwidth]{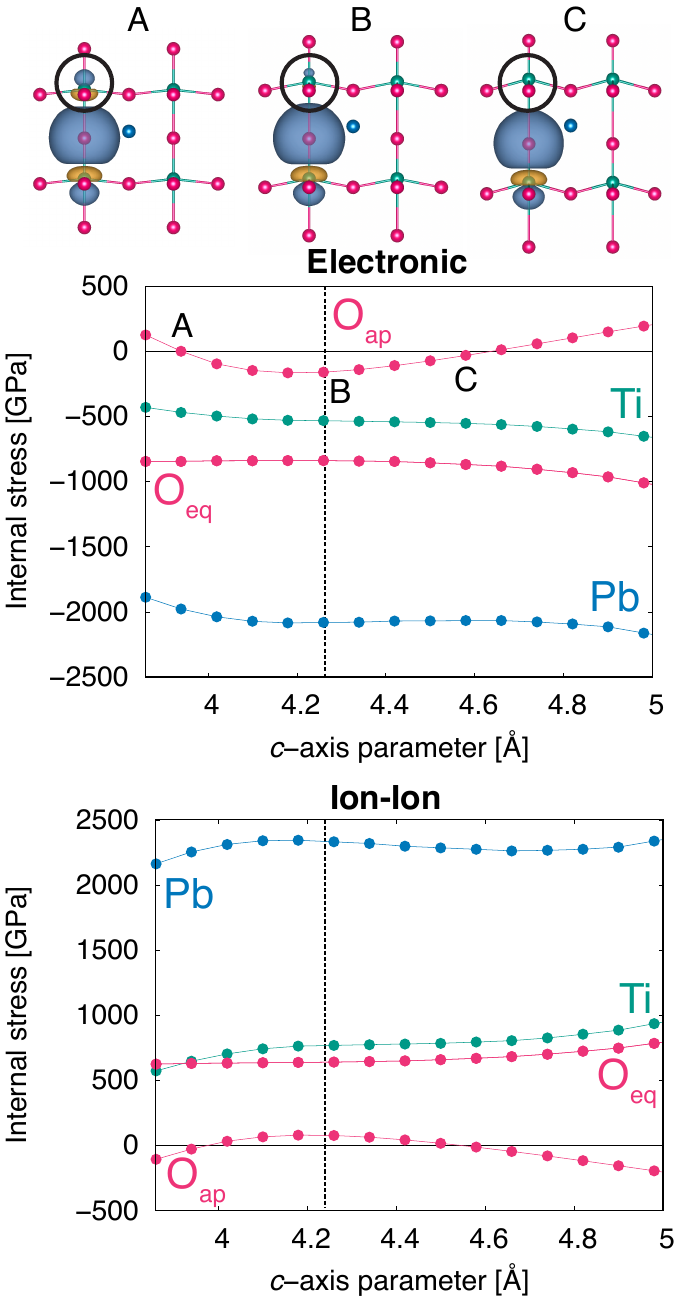}
\caption{
(\textit{Top}) Variations in the orbitally resolved internal stresses (summed over atoms in the primitive cell) felt by the MLWFs associated with \ce{Pb}, \ce{Ti}, \ce{O_{eq}}, and \ce{O_{ap}} in \ce{PbTiO3} with applied stress. To show how the structure is changing with applied stress, the $c$-axis lattice parameters resulting from the applied stress are plotted on the $x$-axis, rather than the applied stress itself. Positions A, B, and C denote the $c$-axis lattice parameters for which the plotted \ce{O_{ap}}-$2s$ MLWF is shown in the top inset. The circled areas on each image are intended to draw the eye to the \ce{Ti} atom, in particular the loss of hybridization with the \ce{O_{ap}}-$2s$ states as the applied stress increases and the $c$-axis lengthens. (\textit{Bottom}) Ionically resolved internal stress associated with $E_{\rm ext}^{\rm (I-I)}$. Vertical lines in each plot correspond to the computed equilibrium $c$-axis lattice parameter for \ce{PbTiO3}. The lines are guides for the eyes.}
\label{fig:MLWF_pto}
\end{figure}
%
%

Fig.~S1 and Fig.~\ref{energies}c show how $E_{\rm tot}$, $E_{\rm elec}$, and $E_{\rm ext}^{\rm (I-I)}$ change as a function of the \emph{applied} stress. However, there is also a corresponding \emph{internal} stress, \ie the force felt by the electrons and ions due to the applied stress~\cite{nielsen85},
\begin{align}
    \sigma_{\alpha\beta}^{\rm int} = \frac{1}{\Omega}\frac{\partial E}{\partial \varepsilon_{\alpha\beta}},
    \label{eq:internal}
\end{align}
where $E$ is an energy (\eg $E_{\rm tot}$, $E_{\rm elec}$, $E_{\rm ext}^{\rm (I-I)}$, or as we will soon see, the energy associated with individual MLWFs) and $\varepsilon_{\alpha\beta}$ is the strain induced by the applied stress along Cartesian directions $\alpha$ and $\beta$ (we define strain as the engineering strain with respect to the fully relaxed equilibrium $c$-axis lattice parameter of \ce{PbTiO3} from our calculations). Fig.~\ref{energies}d shows that the internal stress associated with $E_{\rm elec}$ ($E_{\rm ext}^{\rm (I-I)}$) is negative (positive), \ie $E_{\rm elec}$ ($E_{\rm ext}^{\rm (I-I)}$) decreases (increases) as $\sigma_{zz}$ increases (\cf Fig.~\ref{energies}c and Eq.~\eqref{eq:internal}). 

We now partition the internal stress associated with $E_{\rm elec}$ into orbitally resolved contributions from the MLWFs associated with the Pb ($5s,5p,5d,6s$), Ti ($3s,3p$), \ce{O_{eq}} ($2s,2p$), and \ce{O_{ap}} ($2s,2p$) atoms (we sum the contributions from all MLWFs associated with a given atom; individual MLWF contributions are provided in Fig.~S2~\cite{supp}). Fig.~\ref{energies}b shows an anomalous region near equilibrium. Approaching the anomaly from the left, we observe a decreasing slope. Passing through the anomaly, the slope then increases. If the anomaly arises at least in part from the $E_{\rm elec}$ contribution to the total energy, then close to the anomaly we may see strongly non-linear MLWF contributions to the internal stress through Eq.~\eqref{eq:decomposition}. Each MLWF can feel a different internal stress, akin to a set of parallel springs between two plates. This raises the question, how do the individual MLWFs in \ce{PbTiO3} contribute to the observed deviation from linearity?

As shown in Fig.~\ref{fig:MLWF_pto}, the internal stresses felt by the MLWFs associated with the \ce{Pb}, \ce{Ti}, and \ce{O_{eq}} atoms are effectively flat in the immediate vicinity of the equilibrium $c$-axis lattice parameter, and show negligible curvature with increasing applied stress. 
The only MLWFs that significantly change (contribute to the change in slope shown in Fig.~\ref{energies}b) are the \ce{O_{ap}} ($2s,2p$) MLWFs. A similar trend can also be seen for the ionically resolved $E_{\rm ext}^{\rm (I-I)}$ internal stress in Fig.~\ref{fig:MLWF_pto} (see also the explanatory note in Ref. \onlinecite{e-ion-ion} and Fig. S3 in Ref. \onlinecite{supp}, which shows the atom-resolved contribution to the internal stress summed over both $E_{\rm elec}$ and $E_{\rm ext}^{\rm (I-I)}$). The appearance of two zero-point crossings in the \ce{O_{ap}} MLWF internal stress, at positions A and C in Fig.~\ref{fig:MLWF_pto}, shows that the energy landscape is perched near a barrier connecting two (meta-)stable configurations, \eg a double-well or corrugated potential, corroborating the expected deviation from linearity.

Fig.~\ref{fig:MLWF_pto} also shows how the MLWF associated with the \ce{O_{ap}}-$2s$ state changes as the $c$-axis lattice parameter is altered \via the applied stress. When the $c$-axis lattice parameter is smaller than its equilibrium value (position A, $3.94$~\AA), there is significant hybridization of the \ce{O_{ap}}-$2s$ state with the $d$ orbitals of the neighboring Ti atoms, as is well known from previous work~\cite{cohen92,marzari98}. As the $c$-axis increases, there is a loss of covalency in the \ce{Ti-O_{ap}} bond (which was quantified \via MLWF population analysis~\cite{bhattacharjee2010wannier}, see Ref. \onlinecite{supp}) and a transfer of charge back onto the \ce{Ti} and \ce{O_{ap}} atoms (position B, $4.26$~\AA). As the applied stress increases further, the $d$ character of the orbital associated with one of the neighboring Ti atoms decreases and eventually disappears (position C, $4.66$~\AA).
%
%
\begin{figure}
    \centering
    \includegraphics[width=0.99\columnwidth]{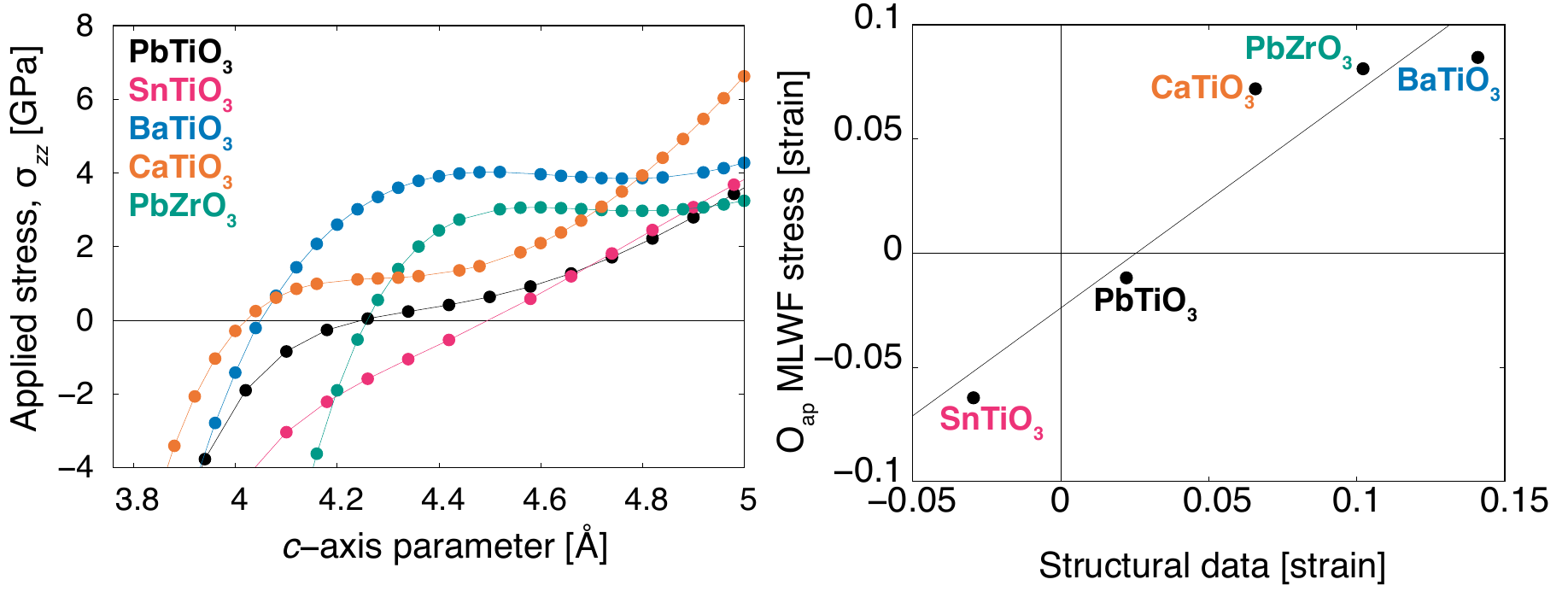}
    \caption{(\textit{Left}) Variation in the $c$-axis lattice parameter of various perovskites in the $P4mm$ space group with applied stress from our first-principles calculations. 
    (\textit{Right}) The strain value of the elastic anomaly compared with its predicted value using MLWF stress alone. The elastic anomaly is defined from structural data as the inflection point of the stress-strain curve, corresponding to the peak compliance, and is predicted from the \ce{O_{ap}} MLWF stress as the (local) minimum. The line is a linear fit to our data and is shown as a guide to the eye.  
    }
    \label{stress_strain_all}
\end{figure}
%
%

Our picture of the elastic anomaly in \ce{PbTiO3} is that it represents a kind of isostructural phase transition \cite{tinte03,moriwake08,stengel09} associated with changes in the \ce{O_{ap}} bonding environment with applied stress. On the left side of the transition is a material with an octahedrally coordinated Ti atom with two \ce{Ti-O_{ap}} bonds along the tetragonal axis containing significant covalency. On the right side of the anomaly is a material where the \ce{Ti} atom is better described as being five-coordinate and no longer as compliant with respect to further applied stress, $\sigma_{zz}$. This interpretation is corroborated by the evolution of the internal stress on the \ce{O_{ap}} states illustrated in Fig.~\ref{fig:MLWF_pto}. These states are under compression at applied stresses within the anomalous region, which causes them to `push outwards' against the walls of the unit cell along the $c$-axis until they reach a new favorable bonding environment. That is, the \ce{O_{ap}} states would prefer an elongated $c$-axis.

Fig.~\ref{stress_strain_all} shows that a number of other perovskites exhibit a similar anomaly under applied stress when placed in the $P4mm$ space group (Refs.~\onlinecite{tinte03} and \onlinecite{moriwake08} also reported results for other perovskites under negative \emph{pressures}). Our results suggest that the \emph{existence} of the elastic anomaly itself appears tied to the presence of a corrugated potential for the \ce{O_{ap}} atom in the tetragonal $P4mm$ space group. The \emph{position} of the anomaly with respect to the equilibrium structure is apparently controlled by the A-site cation.

Of all the materials we considered, the anomaly is closest to the equilibrium structure for \ce{PbTiO3}, and this contributes to the utility of \ce{PbTiO3} as an excellent piezoelectric. Indeed, previous theoretical and experimental work has shown that large enhancements in certain piezoelectric stress and strain coefficients under modest amounts of hydrostatic pressure, rather than stress, are due to sharp decreases in the enthalpy difference between the tetragonal and rhombohedral phases and the formation of a morphotrophic phase boundary~\cite{wu05} (alloying \ce{PbTiO3} with other materials, such as \ce{PbZrO3}, is thought to tune the phase boundary down to ambient pressure~\cite{ahart08}). Our work has revealed the central role of the \ce{O_{ap}} atoms, and changes in their bonding environment, in driving this behavior, an insight that would be difficult to obtain using other techniques. 

We can further strengthen the association between the \ce{O_{ap}} bonding environment and the anomaly, as shown in Fig.~\ref{stress_strain_all}---there is a roughly linear relationship between the location of the anomaly obtained from structural data and the strain that minimizes the total \ce{O_{ap}} MLWF internal stress for the materials investigated here. That is, the technique used here reveals the link between a macroscopic mechanical property (structural response to applied stress) and the microscopic details of particular electronic states. Additional work would be required to better understand this association, and in particular, to elucidate how to tune the \ce{O_{ap}} bonding environment to tailor the elastic response. However, we have shown that the technique presented here can provide an additional channel of information to aid design efforts.
%
%
\begin{figure}
    \centering
    \includegraphics[width=\columnwidth]{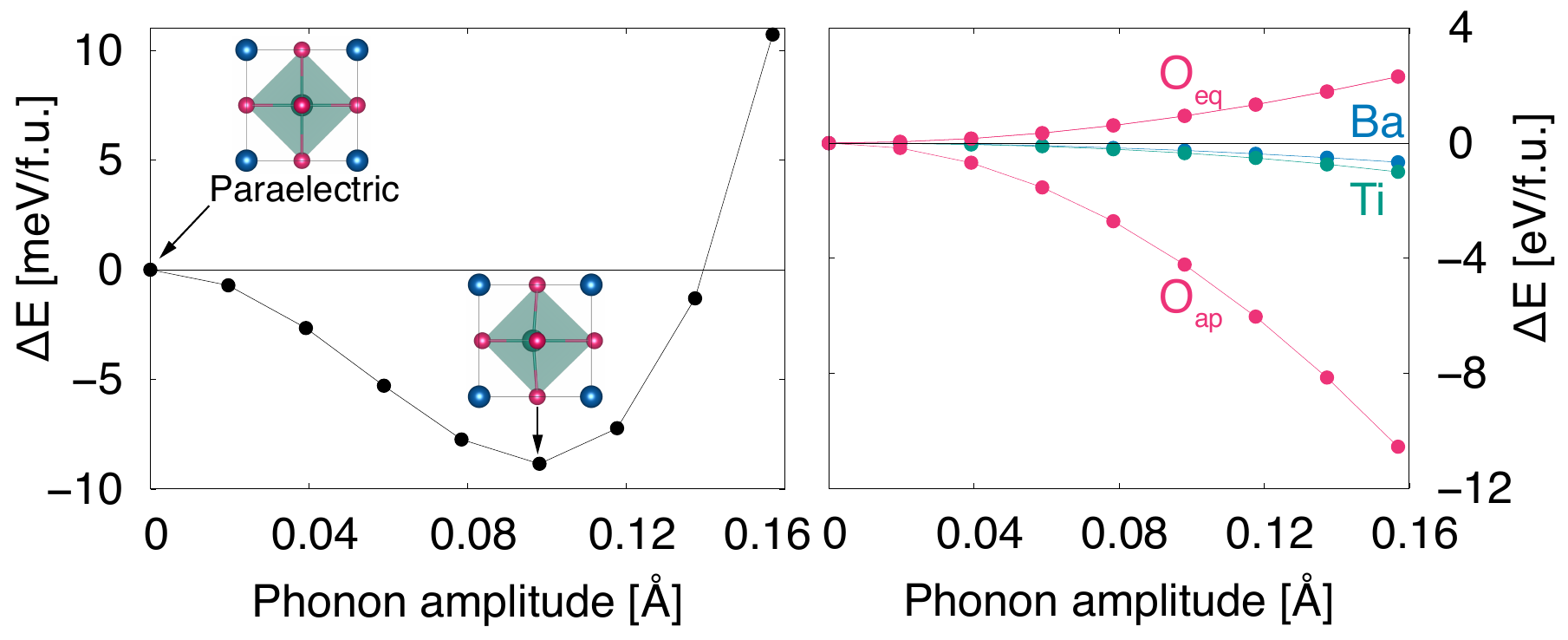}
    \caption{(\textit{Left}) Change in the total energy of cubic paraelectric BaTiO$_3$ ($Pm\bar{3}m$) as a function of the amplitude of the unstable $\Gamma_4^-$ phonon. (\textit{Right}) Change in energy partitioned into contributions from MLWFs centered on the atoms. Note the difference in energy units between the left and right panels. The lines are guides for the eyes.} 
    \label{bto_phonon}
\end{figure}
%
%

\textit{Extension to Other Properties}.\ $\mathrm{-}$ Our technique can also be used to gain chemical insight into more than just mechanical properties. \ce{BaTiO3} undergoes a structural transition to a phase with $P4mm$ symmetry around $400$~K, which is driven by a phonon of the cubic phase transforming like the irreducible representation $\Gamma_4^-$~\cite{zhong94}. When this phonon is `frozen in' to the cubic phase in a semi-local DFT calculation, it gives rise to a spontaneous polarization and lowers the total energy, as shown in Fig.~\ref{bto_phonon}. By partitioning this energy change into contributions from the MLWFs (again, summed over atomic sites), we can see that changes in the \ce{O_{ap}} ($2s,2p$) MLWFs contribute the most to the energy lowering, whereas changes in the (filled) states associated with \ce{Ba} and \ce{Ti} barely lower the energy. This is consistent with the consensus model for this transition, which posits that the primary driver is a change in cross-gap hybridization between the formally filled O-$p$ states and formally empty Ti-$d$ states~\cite{cohen92,marzari98}. However, there are many systems for which no such models exist, making it difficult to elucidate mechanisms for phase transitions. In particular, it is dangerous to assume that the atoms which make the largest displacements in a phase transition are the most important to the mechanism, since those atoms may simply have highly anharmonic local energy surfaces. Our technique works directly with the quantity that really matters---the total energy---and unambiguously partitions it into contributions from different electronic states and ion-ion contributions associated with specific atoms, thereby revealing their respective roles in structural phase transitions of interest. The technique presented here could therefore become a vital tool in model-building efforts for new materials, and for existing systems for which available tools have proven inadequate.

\begin{acknowledgments}
This work was supported by the National Science Foundation under Grant Nos.\ DMR-1550347 (ETR, NAB), DMR-1719875 (GK), and CHE-1945676 (HK, JZ, RAD). NAB and RAD acknowledge a New Directions award (62001-ND10) from the American Chemical Society Petroleum Research Fund. RAD also gratefully acknowledges financial support from an Alfred P.\ Sloan Research Fellowship.
\end{acknowledgments}

\bibliography{references}

\begin{thebibliography}{47}%
\makeatletter
\providecommand \@ifxundefined [1]{%
 \@ifx{#1\undefined}
}%
\providecommand \@ifnum [1]{%
 \ifnum #1\expandafter \@firstoftwo
 \else \expandafter \@secondoftwo
 \fi
}%
\providecommand \@ifx [1]{%
 \ifx #1\expandafter \@firstoftwo
 \else \expandafter \@secondoftwo
 \fi
}%
\providecommand \natexlab [1]{#1}%
\providecommand \enquote  [1]{``#1''}%
\providecommand \bibnamefont  [1]{#1}%
\providecommand \bibfnamefont [1]{#1}%
\providecommand \citenamefont [1]{#1}%
\providecommand \href@noop [0]{\@secondoftwo}%
\providecommand \href [0]{\begingroup \@sanitize@url \@href}%
\providecommand \@href[1]{\@@startlink{#1}\@@href}%
\providecommand \@@href[1]{\endgroup#1\@@endlink}%
\providecommand \@sanitize@url [0]{\catcode `\\12\catcode `\$12\catcode
  `\&12\catcode `\#12\catcode `\^12\catcode `\_12\catcode `\%12\relax}%
\providecommand \@@startlink[1]{}%
\providecommand \@@endlink[0]{}%
\providecommand \url  [0]{\begingroup\@sanitize@url \@url }%
\providecommand \@url [1]{\endgroup\@href {#1}{\urlprefix }}%
\providecommand \urlprefix  [0]{URL }%
\providecommand \Eprint [0]{\href }%
\providecommand \doibase [0]{https://doi.org/}%
\providecommand \selectlanguage [0]{\@gobble}%
\providecommand \bibinfo  [0]{\@secondoftwo}%
\providecommand \bibfield  [0]{\@secondoftwo}%
\providecommand \translation [1]{[#1]}%
\providecommand \BibitemOpen [0]{}%
\providecommand \bibitemStop [0]{}%
\providecommand \bibitemNoStop [0]{.\EOS\space}%
\providecommand \EOS [0]{\spacefactor3000\relax}%
\providecommand \BibitemShut  [1]{\csname bibitem#1\endcsname}%
\let\auto@bib@innerbib\@empty
\bibitem [{\citenamefont {Manthiram}(2020)}]{manthiram20}%
  \BibitemOpen
  \bibfield  {author} {\bibinfo {author} {\bibfnamefont {A.}~\bibnamefont
  {Manthiram}},\ }\bibfield  {title} {\bibinfo {title} {A reflection on
  lithium-ion battery cathode chemistry},\ }\href
  {https://doi.org/10.1038/s41467-020-15355-0} {\bibfield  {journal} {\bibinfo
  {journal} {Nat. Commun.}\ }\textbf {\bibinfo {volume} {11}},\ \bibinfo
  {pages} {1550} (\bibinfo {year} {2020})}\BibitemShut {NoStop}%
\bibitem [{\citenamefont {Burg}\ and\ \citenamefont
  {Dauskardt}(2016)}]{burg16}%
  \BibitemOpen
  \bibfield  {author} {\bibinfo {author} {\bibfnamefont {J.~A.}\ \bibnamefont
  {Burg}}\ and\ \bibinfo {author} {\bibfnamefont {R.~H.}\ \bibnamefont
  {Dauskardt}},\ }\bibfield  {title} {\bibinfo {title} {Elastic and thermal
  expansion asymmetry in dense molecular materials},\ }\href
  {https://doi.org/10.1038/nmat4674} {\bibfield  {journal} {\bibinfo  {journal}
  {Nat. Mater.}\ }\textbf {\bibinfo {volume} {15}},\ \bibinfo {pages} {974}
  (\bibinfo {year} {2016})}\BibitemShut {NoStop}%
\bibitem [{\citenamefont {Mason}\ \emph {et~al.}(2015)\citenamefont {Mason},
  \citenamefont {Oktawiec}, \citenamefont {Taylor}, \citenamefont {Hudson},
  \citenamefont {Rodriguez}, \citenamefont {Bachman}, \citenamefont {Gonzalez},
  \citenamefont {Cervellino}, \citenamefont {Guagliardi}, \citenamefont
  {Brown}, \citenamefont {Llewellyn}, \citenamefont {Masciocchi},\ and\
  \citenamefont {Long}}]{mason15}%
  \BibitemOpen
  \bibfield  {author} {\bibinfo {author} {\bibfnamefont {J.~A.}\ \bibnamefont
  {Mason}}, \bibinfo {author} {\bibfnamefont {J.}~\bibnamefont {Oktawiec}},
  \bibinfo {author} {\bibfnamefont {M.~K.}\ \bibnamefont {Taylor}}, \bibinfo
  {author} {\bibfnamefont {M.~R.}\ \bibnamefont {Hudson}}, \bibinfo {author}
  {\bibfnamefont {J.}~\bibnamefont {Rodriguez}}, \bibinfo {author}
  {\bibfnamefont {J.~E.}\ \bibnamefont {Bachman}}, \bibinfo {author}
  {\bibfnamefont {M.~I.}\ \bibnamefont {Gonzalez}}, \bibinfo {author}
  {\bibfnamefont {A.}~\bibnamefont {Cervellino}}, \bibinfo {author}
  {\bibfnamefont {A.}~\bibnamefont {Guagliardi}}, \bibinfo {author}
  {\bibfnamefont {C.~M.}\ \bibnamefont {Brown}}, \bibinfo {author}
  {\bibfnamefont {P.~L.}\ \bibnamefont {Llewellyn}}, \bibinfo {author}
  {\bibfnamefont {N.}~\bibnamefont {Masciocchi}},\ and\ \bibinfo {author}
  {\bibfnamefont {J.~R.}\ \bibnamefont {Long}},\ }\bibfield  {title} {\bibinfo
  {title} {Methane storage in flexible metal–organic frameworks with
  intrinsic thermal management},\ }\href {https://doi.org/10.1038/nature15732}
  {\bibfield  {journal} {\bibinfo  {journal} {Nature}\ }\textbf {\bibinfo
  {volume} {527}},\ \bibinfo {pages} {357} (\bibinfo {year}
  {2015})}\BibitemShut {NoStop}%
\bibitem [{\citenamefont {{de Jong}}\ \emph {et~al.}(2015)\citenamefont {{de
  Jong}}, \citenamefont {Chen}, \citenamefont {Angsten}, \citenamefont {Jain},
  \citenamefont {Notestine}, \citenamefont {Gamst}, \citenamefont {Sluiter},
  \citenamefont {Krishna~Ande}, \citenamefont {{van der Zwaag}}, \citenamefont
  {Plata}, \citenamefont {Toher}, \citenamefont {Curtarolo}, \citenamefont
  {Ceder}, \citenamefont {Persson},\ and\ \citenamefont {Asta}}]{dejong15}%
  \BibitemOpen
  \bibfield  {author} {\bibinfo {author} {\bibfnamefont {M.}~\bibnamefont {{de
  Jong}}}, \bibinfo {author} {\bibfnamefont {W.}~\bibnamefont {Chen}}, \bibinfo
  {author} {\bibfnamefont {T.}~\bibnamefont {Angsten}}, \bibinfo {author}
  {\bibfnamefont {A.}~\bibnamefont {Jain}}, \bibinfo {author} {\bibfnamefont
  {R.}~\bibnamefont {Notestine}}, \bibinfo {author} {\bibfnamefont
  {A.}~\bibnamefont {Gamst}}, \bibinfo {author} {\bibfnamefont
  {M.}~\bibnamefont {Sluiter}}, \bibinfo {author} {\bibfnamefont
  {C.}~\bibnamefont {Krishna~Ande}}, \bibinfo {author} {\bibfnamefont
  {S.}~\bibnamefont {{van der Zwaag}}}, \bibinfo {author} {\bibfnamefont
  {J.~J.}\ \bibnamefont {Plata}}, \bibinfo {author} {\bibfnamefont
  {C.}~\bibnamefont {Toher}}, \bibinfo {author} {\bibfnamefont
  {S.}~\bibnamefont {Curtarolo}}, \bibinfo {author} {\bibfnamefont
  {G.}~\bibnamefont {Ceder}}, \bibinfo {author} {\bibfnamefont {K.~A.}\
  \bibnamefont {Persson}},\ and\ \bibinfo {author} {\bibfnamefont
  {M.}~\bibnamefont {Asta}},\ }\bibfield  {title} {\bibinfo {title} {Charting
  the complete elastic properties of inorganic crystalline compounds},\ }\href
  {https://doi.org/10.1038/sdata.2015.9} {\bibfield  {journal} {\bibinfo
  {journal} {Sci. Data}\ }\textbf {\bibinfo {volume} {2}},\ \bibinfo {pages}
  {150009} (\bibinfo {year} {2015})}\BibitemShut {NoStop}%
\bibitem [{\citenamefont {Gibbs}\ \emph {et~al.}(1998)\citenamefont {Gibbs},
  \citenamefont {Hill}, \citenamefont {Boisen},\ and\ \citenamefont
  {Downs}}]{gibbs98}%
  \BibitemOpen
  \bibfield  {author} {\bibinfo {author} {\bibfnamefont {G.~V.}\ \bibnamefont
  {Gibbs}}, \bibinfo {author} {\bibfnamefont {F.~C.}\ \bibnamefont {Hill}},
  \bibinfo {author} {\bibfnamefont {M.~B.}\ \bibnamefont {Boisen}},\ and\
  \bibinfo {author} {\bibfnamefont {R.~T.}\ \bibnamefont {Downs}},\ }\bibfield
  {title} {\bibinfo {title} {Power law relationships between bond length, bond
  strength and electron density distributions},\ }\href
  {https://doi.org/10.1007/s002690050151} {\bibfield  {journal} {\bibinfo
  {journal} {Phys. Chem. Miner.}\ }\textbf {\bibinfo {volume} {25}},\ \bibinfo
  {pages} {585} (\bibinfo {year} {1998})}\BibitemShut {NoStop}%
\bibitem [{\citenamefont {Grochala}\ \emph {et~al.}(2007)\citenamefont
  {Grochala}, \citenamefont {Hoffmann}, \citenamefont {Feng},\ and\
  \citenamefont {Ashcroft}}]{grochala07}%
  \BibitemOpen
  \bibfield  {author} {\bibinfo {author} {\bibfnamefont {W.}~\bibnamefont
  {Grochala}}, \bibinfo {author} {\bibfnamefont {R.}~\bibnamefont {Hoffmann}},
  \bibinfo {author} {\bibfnamefont {J.}~\bibnamefont {Feng}},\ and\ \bibinfo
  {author} {\bibfnamefont {N.~W.}\ \bibnamefont {Ashcroft}},\ }\bibfield
  {title} {\bibinfo {title} {The chemical imagination at work in very tight
  places},\ }\href {https://doi.org/10.1002/anie.200602485} {\bibfield
  {journal} {\bibinfo  {journal} {Angew. Chem. Int. Ed.}\ }\textbf {\bibinfo
  {volume} {46}},\ \bibinfo {pages} {3620} (\bibinfo {year}
  {2007})}\BibitemShut {NoStop}%
\bibitem [{\citenamefont {Zeier}\ \emph {et~al.}(2016)\citenamefont {Zeier},
  \citenamefont {Zevalkink}, \citenamefont {Gibbs}, \citenamefont {Hautier},
  \citenamefont {Kanatzidis},\ and\ \citenamefont {Snyder}}]{zeier16}%
  \BibitemOpen
  \bibfield  {author} {\bibinfo {author} {\bibfnamefont {W.~G.}\ \bibnamefont
  {Zeier}}, \bibinfo {author} {\bibfnamefont {A.}~\bibnamefont {Zevalkink}},
  \bibinfo {author} {\bibfnamefont {Z.~M.}\ \bibnamefont {Gibbs}}, \bibinfo
  {author} {\bibfnamefont {G.}~\bibnamefont {Hautier}}, \bibinfo {author}
  {\bibfnamefont {M.~G.}\ \bibnamefont {Kanatzidis}},\ and\ \bibinfo {author}
  {\bibfnamefont {G.~J.}\ \bibnamefont {Snyder}},\ }\bibfield  {title}
  {\bibinfo {title} {Thinking like a chemist: {Intuition} in thermoelectric
  materials},\ }\href {https://doi.org/10.1002/anie.201508381} {\bibfield
  {journal} {\bibinfo  {journal} {Angew. Chem. Int. Ed.}\ }\textbf {\bibinfo
  {volume} {55}},\ \bibinfo {pages} {6826} (\bibinfo {year}
  {2016})}\BibitemShut {NoStop}%
\bibitem [{\citenamefont {Cohen}(1985)}]{cohen85}%
  \BibitemOpen
  \bibfield  {author} {\bibinfo {author} {\bibfnamefont {M.~L.}\ \bibnamefont
  {Cohen}},\ }\bibfield  {title} {\bibinfo {title} {Calculation of bulk moduli
  of diamond and zinc-blende solids},\ }\href
  {https://doi.org/10.1103/PhysRevB.32.7988} {\bibfield  {journal} {\bibinfo
  {journal} {Phys. Rev. B}\ }\textbf {\bibinfo {volume} {32}},\ \bibinfo
  {pages} {7988} (\bibinfo {year} {1985})}\BibitemShut {NoStop}%
\bibitem [{\citenamefont {Harrison}(2004)}]{harrison}%
  \BibitemOpen
  \bibfield  {author} {\bibinfo {author} {\bibfnamefont {W.~A.}\ \bibnamefont
  {Harrison}},\ }\href {https://doi.org/10.1142/5432} {\emph {\bibinfo {title}
  {Elementary Electronic Structure}}},\ \bibinfo {edition} {{Revised}}\ ed.\
  (\bibinfo  {publisher} {World Scientific},\ \bibinfo {address} {Singapore},\
  \bibinfo {year} {2004})\BibitemShut {NoStop}%
\bibitem [{\citenamefont {Marzari}\ and\ \citenamefont
  {Vanderbilt}(1997)}]{marzari97}%
  \BibitemOpen
  \bibfield  {author} {\bibinfo {author} {\bibfnamefont {N.}~\bibnamefont
  {Marzari}}\ and\ \bibinfo {author} {\bibfnamefont {D.}~\bibnamefont
  {Vanderbilt}},\ }\bibfield  {title} {\bibinfo {title} {Maximally localized
  generalized {Wannier} functions for composite energy bands},\ }\href
  {https://doi.org/10.1103/PhysRevB.56.12847} {\bibfield  {journal} {\bibinfo
  {journal} {Phys. Rev. B}\ }\textbf {\bibinfo {volume} {56}},\ \bibinfo
  {pages} {12847} (\bibinfo {year} {1997})}\BibitemShut {NoStop}%
\bibitem [{\citenamefont {Marzari}\ \emph {et~al.}(2012)\citenamefont
  {Marzari}, \citenamefont {Mostofi}, \citenamefont {Yates}, \citenamefont
  {Souza},\ and\ \citenamefont {Vanderbilt}}]{marzari12}%
  \BibitemOpen
  \bibfield  {author} {\bibinfo {author} {\bibfnamefont {N.}~\bibnamefont
  {Marzari}}, \bibinfo {author} {\bibfnamefont {A.~A.}\ \bibnamefont
  {Mostofi}}, \bibinfo {author} {\bibfnamefont {J.~R.}\ \bibnamefont {Yates}},
  \bibinfo {author} {\bibfnamefont {I.}~\bibnamefont {Souza}},\ and\ \bibinfo
  {author} {\bibfnamefont {D.}~\bibnamefont {Vanderbilt}},\ }\bibfield  {title}
  {\bibinfo {title} {Maximally localized {Wannier} functions: {Theory} and
  applications},\ }\href {https://doi.org/10.1103/RevModPhys.84.1419}
  {\bibfield  {journal} {\bibinfo  {journal} {Rev. Mod. Phys.}\ }\textbf
  {\bibinfo {volume} {84}},\ \bibinfo {pages} {1419} (\bibinfo {year}
  {2012})}\BibitemShut {NoStop}%
\bibitem [{\citenamefont {Majewski}\ and\ \citenamefont
  {Vogl}(1986)}]{majewski1986crystal}%
  \BibitemOpen
  \bibfield  {author} {\bibinfo {author} {\bibfnamefont {J.~A.}\ \bibnamefont
  {Majewski}}\ and\ \bibinfo {author} {\bibfnamefont {P.}~\bibnamefont
  {Vogl}},\ }\bibfield  {title} {\bibinfo {title} {Crystal stability and
  structural transition pressures of $sp$-bonded solids},\ }\href
  {https://doi.org/10.1103/PhysRevLett.57.1366} {\bibfield  {journal} {\bibinfo
   {journal} {Phys. Rev. Lett.}\ }\textbf {\bibinfo {volume} {57}},\ \bibinfo
  {pages} {1366} (\bibinfo {year} {1986})}\BibitemShut {NoStop}%
\bibitem [{\citenamefont {Shirane}\ and\ \citenamefont
  {Hoshino}(1951)}]{shirane1951phase}%
  \BibitemOpen
  \bibfield  {author} {\bibinfo {author} {\bibfnamefont {G.}~\bibnamefont
  {Shirane}}\ and\ \bibinfo {author} {\bibfnamefont {S.}~\bibnamefont
  {Hoshino}},\ }\bibfield  {title} {\bibinfo {title} {On the phase transition
  in lead titanate},\ }\href {https://doi.org/10.1143/JPSJ.6.265} {\bibfield
  {journal} {\bibinfo  {journal} {J. Phys. Soc. Jpn.}\ }\textbf {\bibinfo
  {volume} {6}},\ \bibinfo {pages} {265} (\bibinfo {year} {1951})}\BibitemShut
  {NoStop}%
\bibitem [{\citenamefont {Chen}\ \emph {et~al.}(2005)\citenamefont {Chen},
  \citenamefont {Xing}, \citenamefont {Yu},\ and\ \citenamefont
  {Liu}}]{chen2005structure}%
  \BibitemOpen
  \bibfield  {author} {\bibinfo {author} {\bibfnamefont {J.}~\bibnamefont
  {Chen}}, \bibinfo {author} {\bibfnamefont {X.~R.}\ \bibnamefont {Xing}},
  \bibinfo {author} {\bibfnamefont {R.~B.}\ \bibnamefont {Yu}},\ and\ \bibinfo
  {author} {\bibfnamefont {G.~R.}\ \bibnamefont {Liu}},\ }\bibfield  {title}
  {\bibinfo {title} {Structure and enhancement of negative thermal expansion in
  the {PbTiO$_3$}–{CdTiO$_3$} system},\ }\href
  {https://doi.org/10.1063/1.2140486} {\bibfield  {journal} {\bibinfo
  {journal} {Appl. Phys. Lett.}\ }\textbf {\bibinfo {volume} {87}},\ \bibinfo
  {pages} {231915} (\bibinfo {year} {2005})}\BibitemShut {NoStop}%
\bibitem [{\citenamefont {Ritz}\ and\ \citenamefont
  {Benedek}(2018)}]{ritz2018}%
  \BibitemOpen
  \bibfield  {author} {\bibinfo {author} {\bibfnamefont {E.~T.}\ \bibnamefont
  {Ritz}}\ and\ \bibinfo {author} {\bibfnamefont {N.~A.}\ \bibnamefont
  {Benedek}},\ }\bibfield  {title} {\bibinfo {title} {Interplay between phonons
  and anisotropic elasticity drives negative thermal expansion in
  {PbTiO$_3$}},\ }\href {https://doi.org/10.1103/PhysRevLett.121.255901}
  {\bibfield  {journal} {\bibinfo  {journal} {Phys. Rev. Lett.}\ }\textbf
  {\bibinfo {volume} {121}},\ \bibinfo {pages} {255901} (\bibinfo {year}
  {2018})}\BibitemShut {NoStop}%
\bibitem [{\citenamefont {Tinte}\ \emph {et~al.}(2003)\citenamefont {Tinte},
  \citenamefont {Rabe},\ and\ \citenamefont {Vanderbilt}}]{tinte03}%
  \BibitemOpen
  \bibfield  {author} {\bibinfo {author} {\bibfnamefont {S.}~\bibnamefont
  {Tinte}}, \bibinfo {author} {\bibfnamefont {K.~M.}\ \bibnamefont {Rabe}},\
  and\ \bibinfo {author} {\bibfnamefont {D.}~\bibnamefont {Vanderbilt}},\
  }\bibfield  {title} {\bibinfo {title} {Anomalous enhancement of tetragonality
  in {PbTiO$_3$} induced by negative pressure},\ }\href
  {https://doi.org/10.1103/PhysRevB.68.144105} {\bibfield  {journal} {\bibinfo
  {journal} {Phys. Rev. B}\ }\textbf {\bibinfo {volume} {68}},\ \bibinfo
  {pages} {144105} (\bibinfo {year} {2003})}\BibitemShut {NoStop}%
\bibitem [{\citenamefont {Duan}\ \emph {et~al.}(2008)\citenamefont {Duan},
  \citenamefont {Shi},\ and\ \citenamefont {Qin}}]{duan08}%
  \BibitemOpen
  \bibfield  {author} {\bibinfo {author} {\bibfnamefont {Y.}~\bibnamefont
  {Duan}}, \bibinfo {author} {\bibfnamefont {H.}~\bibnamefont {Shi}},\ and\
  \bibinfo {author} {\bibfnamefont {L.}~\bibnamefont {Qin}},\ }\bibfield
  {title} {\bibinfo {title} {Studies of tetragonal {PbTiO$_3$} subjected to
  uniaxial stress along the {$c$-axis}},\ }\href
  {https://doi.org/10.1088/0953-8984/20/17/175210} {\bibfield  {journal}
  {\bibinfo  {journal} {J. Phys. Condens. Matter}\ }\textbf {\bibinfo {volume}
  {20}},\ \bibinfo {pages} {175210} (\bibinfo {year} {2008})}\BibitemShut
  {NoStop}%
\bibitem [{\citenamefont {Moriwake}\ \emph {et~al.}(2008)\citenamefont
  {Moriwake}, \citenamefont {Koyama}, \citenamefont {Matsunaga}, \citenamefont
  {Hirayama},\ and\ \citenamefont {Tanaka}}]{moriwake08}%
  \BibitemOpen
  \bibfield  {author} {\bibinfo {author} {\bibfnamefont {H.}~\bibnamefont
  {Moriwake}}, \bibinfo {author} {\bibfnamefont {Y.}~\bibnamefont {Koyama}},
  \bibinfo {author} {\bibfnamefont {K.}~\bibnamefont {Matsunaga}}, \bibinfo
  {author} {\bibfnamefont {T.}~\bibnamefont {Hirayama}},\ and\ \bibinfo
  {author} {\bibfnamefont {I.}~\bibnamefont {Tanaka}},\ }\bibfield  {title}
  {\bibinfo {title} {Isostructural phase transitions of tetragonal perovskite
  titanates under negative hydrostatic pressure},\ }\href
  {https://doi.org/10.1088/0953-8984/20/34/345207} {\bibfield  {journal}
  {\bibinfo  {journal} {J. Phys.: Condens. Matter}\ }\textbf {\bibinfo {volume}
  {20}},\ \bibinfo {pages} {345207} (\bibinfo {year} {2008})}\BibitemShut
  {NoStop}%
\bibitem [{\citenamefont {Sharma}\ \emph {et~al.}(2014)\citenamefont {Sharma},
  \citenamefont {Kreisel},\ and\ \citenamefont {Ghosez}}]{sharma14}%
  \BibitemOpen
  \bibfield  {author} {\bibinfo {author} {\bibfnamefont {H.}~\bibnamefont
  {Sharma}}, \bibinfo {author} {\bibfnamefont {J.}~\bibnamefont {Kreisel}},\
  and\ \bibinfo {author} {\bibfnamefont {P.}~\bibnamefont {Ghosez}},\
  }\bibfield  {title} {\bibinfo {title} {First-principles study of {PbTiO$_3$}
  under uniaxial strains and stresses},\ }\href
  {https://doi.org/10.1103/PhysRevB.90.214102} {\bibfield  {journal} {\bibinfo
  {journal} {Phys. Rev. B}\ }\textbf {\bibinfo {volume} {90}},\ \bibinfo
  {pages} {214102} (\bibinfo {year} {2014})}\BibitemShut {NoStop}%
\bibitem [{\citenamefont {Kvasov}\ \emph {et~al.}(2016)\citenamefont {Kvasov},
  \citenamefont {McGilly}, \citenamefont {Wang}, \citenamefont {Shi},
  \citenamefont {Sandu}, \citenamefont {Sluka}, \citenamefont {Tagantsev},\
  and\ \citenamefont {Setter}}]{setter16}%
  \BibitemOpen
  \bibfield  {author} {\bibinfo {author} {\bibfnamefont {A.}~\bibnamefont
  {Kvasov}}, \bibinfo {author} {\bibfnamefont {L.~J.}\ \bibnamefont {McGilly}},
  \bibinfo {author} {\bibfnamefont {J.}~\bibnamefont {Wang}}, \bibinfo {author}
  {\bibfnamefont {Z.}~\bibnamefont {Shi}}, \bibinfo {author} {\bibfnamefont
  {C.~S.}\ \bibnamefont {Sandu}}, \bibinfo {author} {\bibfnamefont
  {T.}~\bibnamefont {Sluka}}, \bibinfo {author} {\bibfnamefont {A.~K.}\
  \bibnamefont {Tagantsev}},\ and\ \bibinfo {author} {\bibfnamefont
  {N.}~\bibnamefont {Setter}},\ }\bibfield  {title} {\bibinfo {title}
  {Piezoelectric enhancement under negative pressure},\ }\href
  {https://doi.org/10.1038/ncomms12136} {\bibfield  {journal} {\bibinfo
  {journal} {Nat. Commun.}\ }\textbf {\bibinfo {volume} {7}},\ \bibinfo {pages}
  {12136} (\bibinfo {year} {2016})}\BibitemShut {NoStop}%
\bibitem [{\citenamefont {Rahm}\ and\ \citenamefont
  {Hoffmann}(2016)}]{rahm2016distinguishing}%
  \BibitemOpen
  \bibfield  {author} {\bibinfo {author} {\bibfnamefont {M.}~\bibnamefont
  {Rahm}}\ and\ \bibinfo {author} {\bibfnamefont {R.}~\bibnamefont
  {Hoffmann}},\ }\bibfield  {title} {\bibinfo {title} {Distinguishing bonds},\
  }\href {https://doi.org/10.1021/jacs.5b12434} {\bibfield  {journal} {\bibinfo
   {journal} {J. Am. Chem. Soc.}\ }\textbf {\bibinfo {volume} {138}},\ \bibinfo
  {pages} {3731} (\bibinfo {year} {2016})}\BibitemShut {NoStop}%
\bibitem [{\citenamefont {Baba}\ \emph {et~al.}(2006)\citenamefont {Baba},
  \citenamefont {Takeuchi},\ and\ \citenamefont {Nakai}}]{baba2006natural}%
  \BibitemOpen
  \bibfield  {author} {\bibinfo {author} {\bibfnamefont {T.}~\bibnamefont
  {Baba}}, \bibinfo {author} {\bibfnamefont {M.}~\bibnamefont {Takeuchi}},\
  and\ \bibinfo {author} {\bibfnamefont {H.}~\bibnamefont {Nakai}},\ }\bibfield
   {title} {\bibinfo {title} {Natural atomic orbital based energy density
  analysis: {Implementation} and applications},\ }\href
  {https://doi.org/10.1016/j.cplett.2006.03.098} {\bibfield  {journal}
  {\bibinfo  {journal} {Chem. Phys. Lett.}\ }\textbf {\bibinfo {volume}
  {424}},\ \bibinfo {pages} {193} (\bibinfo {year} {2006})}\BibitemShut
  {NoStop}%
\bibitem [{\citenamefont {Maintz}\ \emph {et~al.}(2016)\citenamefont {Maintz},
  \citenamefont {Deringer}, \citenamefont {Tchougr{\'e}eff},\ and\
  \citenamefont {Dronskowski}}]{maintz2016lobster}%
  \BibitemOpen
  \bibfield  {author} {\bibinfo {author} {\bibfnamefont {S.}~\bibnamefont
  {Maintz}}, \bibinfo {author} {\bibfnamefont {V.~L.}\ \bibnamefont
  {Deringer}}, \bibinfo {author} {\bibfnamefont {A.~L.}\ \bibnamefont
  {Tchougr{\'e}eff}},\ and\ \bibinfo {author} {\bibfnamefont {R.}~\bibnamefont
  {Dronskowski}},\ }\bibfield  {title} {\bibinfo {title} {{LOBSTER}: {A} tool
  to extract chemical bonding from plane‐wave based {DFT}},\ }\href
  {https://doi.org/10.1002/jcc.24300} {\bibfield  {journal} {\bibinfo
  {journal} {J. Comput. Chem.}\ }\textbf {\bibinfo {volume} {37}},\ \bibinfo
  {pages} {1030} (\bibinfo {year} {2016})}\BibitemShut {NoStop}%
\bibitem [{\citenamefont {Oliphant}\ \emph {et~al.}(2025)\citenamefont
  {Oliphant}, \citenamefont {Mantena}, \citenamefont {Brod}, \citenamefont
  {Snyder},\ and\ \citenamefont {Sun}}]{oliphant25}%
  \BibitemOpen
  \bibfield  {author} {\bibinfo {author} {\bibfnamefont {E.}~\bibnamefont
  {Oliphant}}, \bibinfo {author} {\bibfnamefont {V.}~\bibnamefont {Mantena}},
  \bibinfo {author} {\bibfnamefont {M.}~\bibnamefont {Brod}}, \bibinfo {author}
  {\bibfnamefont {G.~J.}\ \bibnamefont {Snyder}},\ and\ \bibinfo {author}
  {\bibfnamefont {W.}~\bibnamefont {Sun}},\ }\bibfield  {title} {\bibinfo
  {title} {Why does silicon have an indirect band gap?},\ }\href
  {https://doi.org/10.1039/D4MH01038H} {\bibfield  {journal} {\bibinfo
  {journal} {Mater. Horiz.}\ }\textbf {\bibinfo {volume} {12}},\ \bibinfo
  {pages} {3073} (\bibinfo {year} {2025})}\BibitemShut {NoStop}%
\bibitem [{\citenamefont {Cohen}\ \emph {et~al.}(2000)\citenamefont {Cohen},
  \citenamefont {Frydel}, \citenamefont {Burke},\ and\ \citenamefont
  {Engel}}]{cohen2000total}%
  \BibitemOpen
  \bibfield  {author} {\bibinfo {author} {\bibfnamefont {M.~H.}\ \bibnamefont
  {Cohen}}, \bibinfo {author} {\bibfnamefont {D.}~\bibnamefont {Frydel}},
  \bibinfo {author} {\bibfnamefont {K.}~\bibnamefont {Burke}},\ and\ \bibinfo
  {author} {\bibfnamefont {E.}~\bibnamefont {Engel}},\ }\bibfield  {title}
  {\bibinfo {title} {Total energy density as an interpretative tool},\ }\href
  {https://doi.org/10.1063/1.1286805} {\bibfield  {journal} {\bibinfo
  {journal} {J. Chem. Phys.}\ }\textbf {\bibinfo {volume} {113}},\ \bibinfo
  {pages} {2990} (\bibinfo {year} {2000})}\BibitemShut {NoStop}%
\bibitem [{\citenamefont {Nakai}\ \emph {et~al.}(2007)\citenamefont {Nakai},
  \citenamefont {Kurabayashi}, \citenamefont {Katouda},\ and\ \citenamefont
  {Atsumi}}]{nakai2007extension}%
  \BibitemOpen
  \bibfield  {author} {\bibinfo {author} {\bibfnamefont {H.}~\bibnamefont
  {Nakai}}, \bibinfo {author} {\bibfnamefont {Y.}~\bibnamefont {Kurabayashi}},
  \bibinfo {author} {\bibfnamefont {M.}~\bibnamefont {Katouda}},\ and\ \bibinfo
  {author} {\bibfnamefont {T.}~\bibnamefont {Atsumi}},\ }\bibfield  {title}
  {\bibinfo {title} {Extension of energy density analysis to periodic boundary
  condition calculation: {Evaluation} of locality in extended systems},\ }\href
  {https://doi.org/10.1016/j.cplett.2007.02.054} {\bibfield  {journal}
  {\bibinfo  {journal} {Chem. Phys. Lett.}\ }\textbf {\bibinfo {volume}
  {438}},\ \bibinfo {pages} {132} (\bibinfo {year} {2007})}\BibitemShut
  {NoStop}%
\bibitem [{\citenamefont {Cohen}\ \emph {et~al.}(1994)\citenamefont {Cohen},
  \citenamefont {Mehl},\ and\ \citenamefont
  {Papaconstantopoulos}}]{cohen1994tight}%
  \BibitemOpen
  \bibfield  {author} {\bibinfo {author} {\bibfnamefont {R.~E.}\ \bibnamefont
  {Cohen}}, \bibinfo {author} {\bibfnamefont {M.~J.}\ \bibnamefont {Mehl}},\
  and\ \bibinfo {author} {\bibfnamefont {D.~A.}\ \bibnamefont
  {Papaconstantopoulos}},\ }\bibfield  {title} {\bibinfo {title} {Tight-binding
  total-energy method for transition and noble metals},\ }\href
  {https://doi.org/10.1103/PhysRevB.50.14694} {\bibfield  {journal} {\bibinfo
  {journal} {Phys. Rev. B}\ }\textbf {\bibinfo {volume} {50}},\ \bibinfo
  {pages} {14694} (\bibinfo {year} {1994})}\BibitemShut {NoStop}%
\bibitem [{\citenamefont {Finnis}(2007)}]{finnis2007bond}%
  \BibitemOpen
  \bibfield  {author} {\bibinfo {author} {\bibfnamefont {M.~W.}\ \bibnamefont
  {Finnis}},\ }\bibfield  {title} {\bibinfo {title} {Bond-order potentials
  through the ages},\ }\href {https://doi.org/10.1016/j.pmatsci.2006.10.003}
  {\bibfield  {journal} {\bibinfo  {journal} {Prog. Mater. Sci.}\ }\textbf
  {\bibinfo {volume} {52}},\ \bibinfo {pages} {133} (\bibinfo {year}
  {2007})}\BibitemShut {NoStop}%
\bibitem [{\citenamefont {Pickett}(1989)}]{Pickett1989}%
  \BibitemOpen
  \bibfield  {author} {\bibinfo {author} {\bibfnamefont {W.~E.}\ \bibnamefont
  {Pickett}},\ }\bibfield  {title} {\bibinfo {title} {Pseudopotential methods
  in condensed matter applications},\ }\href
  {https://doi.org/10.1016/0167-7977(89)90002-6} {\bibfield  {journal}
  {\bibinfo  {journal} {Comput. Phys. Rep.}\ }\textbf {\bibinfo {volume} {9}},\
  \bibinfo {pages} {115} (\bibinfo {year} {1989})}\BibitemShut {NoStop}%
\bibitem [{\citenamefont {Schwerdtfeger}(2011)}]{psp_review2011}%
  \BibitemOpen
  \bibfield  {author} {\bibinfo {author} {\bibfnamefont {P.}~\bibnamefont
  {Schwerdtfeger}},\ }\bibfield  {title} {\bibinfo {title} {The pseudopotential
  approximation in electronic structure theory},\ }\href
  {https://doi.org/10.1002/cphc.201100387} {\bibfield  {journal} {\bibinfo
  {journal} {ChemPhysChem}\ }\textbf {\bibinfo {volume} {12}},\ \bibinfo
  {pages} {3143} (\bibinfo {year} {2011})}\BibitemShut {NoStop}%
\bibitem [{sup()}]{supp}%
  \BibitemOpen
  \href@noop {} {}\bibinfo {note} {{See Supplemental Material at [URL will be
  inserted by publisher] for more detailed derivations and computational
  experiments.}}\BibitemShut {Stop}%
\bibitem [{\citenamefont {de~Leeuw}\ \emph {et~al.}(1980)\citenamefont
  {de~Leeuw}, \citenamefont {Perram},\ and\ \citenamefont
  {Smith}}]{de_Leeuw1980}%
  \BibitemOpen
  \bibfield  {author} {\bibinfo {author} {\bibfnamefont {S.~W.}\ \bibnamefont
  {de~Leeuw}}, \bibinfo {author} {\bibfnamefont {J.~W.}\ \bibnamefont
  {Perram}},\ and\ \bibinfo {author} {\bibfnamefont {E.~R.}\ \bibnamefont
  {Smith}},\ }\bibfield  {title} {\bibinfo {title} {Simulation of electrostatic
  systems in periodic boundary conditions. {I}. {L}attice sums and dielectric
  constants},\ }\href {https://doi.org/10.1098/rspa.1980.0135} {\bibfield
  {journal} {\bibinfo  {journal} {Proc. R. Soc. A}\ }\textbf {\bibinfo {volume}
  {373}},\ \bibinfo {pages} {27} (\bibinfo {year} {1980})}\BibitemShut
  {NoStop}%
\bibitem [{\citenamefont {Makov}\ and\ \citenamefont
  {Payne}(1995)}]{Makov1995}%
  \BibitemOpen
  \bibfield  {author} {\bibinfo {author} {\bibfnamefont {G.}~\bibnamefont
  {Makov}}\ and\ \bibinfo {author} {\bibfnamefont {M.~C.}\ \bibnamefont
  {Payne}},\ }\bibfield  {title} {\bibinfo {title} {Periodic boundary
  conditions in \textit{ab initio} calculations},\ }\href
  {https://doi.org/10.1103/PhysRevB.51.4014} {\bibfield  {journal} {\bibinfo
  {journal} {Phys. Rev. B}\ }\textbf {\bibinfo {volume} {51}},\ \bibinfo
  {pages} {4014} (\bibinfo {year} {1995})}\BibitemShut {NoStop}%
\bibitem [{\citenamefont {Giannozzi}\ \emph {et~al.}(2009)\citenamefont
  {Giannozzi}, \citenamefont {Baroni}, \citenamefont {Bonini}, \citenamefont
  {Calandra}, \citenamefont {Car}, \citenamefont {Cavazzoni}, \citenamefont
  {Ceresoli}, \citenamefont {Chiarotti}, \citenamefont {Cococcioni},
  \citenamefont {Dabo}, \citenamefont {{Dal Corso}}, \citenamefont {{de
  Gironcoli}}, \citenamefont {Fabris}, \citenamefont {Fratesi}, \citenamefont
  {Gebauer}, \citenamefont {Gerstmann}, \citenamefont {Gougoussis},
  \citenamefont {Kokalj}, \citenamefont {Lazzeri}, \citenamefont
  {Martin-Samos}, \citenamefont {Marzari}, \citenamefont {Mauri}, \citenamefont
  {Mazzarello}, \citenamefont {Paolini}, \citenamefont {Pasquarello},
  \citenamefont {Paulatto}, \citenamefont {Sbraccia}, \citenamefont {Scandolo},
  \citenamefont {Sclauzero}, \citenamefont {Seitsonen}, \citenamefont
  {Smogunov}, \citenamefont {Umari},\ and\ \citenamefont
  {Wentzcovitch}}]{giannozzi2009quantum}%
  \BibitemOpen
  \bibfield  {author} {\bibinfo {author} {\bibfnamefont {P.}~\bibnamefont
  {Giannozzi}}, \bibinfo {author} {\bibfnamefont {S.}~\bibnamefont {Baroni}},
  \bibinfo {author} {\bibfnamefont {N.}~\bibnamefont {Bonini}}, \bibinfo
  {author} {\bibfnamefont {M.}~\bibnamefont {Calandra}}, \bibinfo {author}
  {\bibfnamefont {R.}~\bibnamefont {Car}}, \bibinfo {author} {\bibfnamefont
  {C.}~\bibnamefont {Cavazzoni}}, \bibinfo {author} {\bibfnamefont
  {D.}~\bibnamefont {Ceresoli}}, \bibinfo {author} {\bibfnamefont {G.~L.}\
  \bibnamefont {Chiarotti}}, \bibinfo {author} {\bibfnamefont {M.}~\bibnamefont
  {Cococcioni}}, \bibinfo {author} {\bibfnamefont {I.}~\bibnamefont {Dabo}},
  \bibinfo {author} {\bibfnamefont {A.}~\bibnamefont {{Dal Corso}}}, \bibinfo
  {author} {\bibfnamefont {S.}~\bibnamefont {{de Gironcoli}}}, \bibinfo
  {author} {\bibfnamefont {S.}~\bibnamefont {Fabris}}, \bibinfo {author}
  {\bibfnamefont {G.}~\bibnamefont {Fratesi}}, \bibinfo {author} {\bibfnamefont
  {R.}~\bibnamefont {Gebauer}}, \bibinfo {author} {\bibfnamefont
  {U.}~\bibnamefont {Gerstmann}}, \bibinfo {author} {\bibfnamefont
  {C.}~\bibnamefont {Gougoussis}}, \bibinfo {author} {\bibfnamefont
  {A.}~\bibnamefont {Kokalj}}, \bibinfo {author} {\bibfnamefont
  {M.}~\bibnamefont {Lazzeri}}, \bibinfo {author} {\bibfnamefont
  {L.}~\bibnamefont {Martin-Samos}}, \bibinfo {author} {\bibfnamefont
  {N.}~\bibnamefont {Marzari}}, \bibinfo {author} {\bibfnamefont
  {F.}~\bibnamefont {Mauri}}, \bibinfo {author} {\bibfnamefont
  {R.}~\bibnamefont {Mazzarello}}, \bibinfo {author} {\bibfnamefont
  {S.}~\bibnamefont {Paolini}}, \bibinfo {author} {\bibfnamefont
  {A.}~\bibnamefont {Pasquarello}}, \bibinfo {author} {\bibfnamefont
  {L.}~\bibnamefont {Paulatto}}, \bibinfo {author} {\bibfnamefont
  {C.}~\bibnamefont {Sbraccia}}, \bibinfo {author} {\bibfnamefont
  {S.}~\bibnamefont {Scandolo}}, \bibinfo {author} {\bibfnamefont
  {G.}~\bibnamefont {Sclauzero}}, \bibinfo {author} {\bibfnamefont {A.~P.}\
  \bibnamefont {Seitsonen}}, \bibinfo {author} {\bibfnamefont {A.}~\bibnamefont
  {Smogunov}}, \bibinfo {author} {\bibfnamefont {P.}~\bibnamefont {Umari}},\
  and\ \bibinfo {author} {\bibfnamefont {R.~M.}\ \bibnamefont {Wentzcovitch}},\
  }\bibfield  {title} {\bibinfo {title} {{Quantum ESPRESSO}: {A} modular and
  open-source software project for quantum simulations of materials},\ }\href
  {https://doi.org/10.1088/0953-8984/21/39/395502} {\bibfield  {journal}
  {\bibinfo  {journal} {J. Phys. Condens. Matter}\ }\textbf {\bibinfo {volume}
  {21}},\ \bibinfo {pages} {395502} (\bibinfo {year} {2009})}\BibitemShut
  {NoStop}%
\bibitem [{\citenamefont {Giannozzi}\ \emph {et~al.}(2017)\citenamefont
  {Giannozzi}, \citenamefont {Andreussi}, \citenamefont {Brumme}, \citenamefont
  {Bunau}, \citenamefont {{Buongiorno Nardelli}}, \citenamefont {Calandra},
  \citenamefont {Car}, \citenamefont {Cavazzoni}, \citenamefont {Ceresoli},
  \citenamefont {Cococcioni}, \citenamefont {Colonna}, \citenamefont
  {Carnimeo}, \citenamefont {{Dal Corso}}, \citenamefont {{de Gironcoli}},
  \citenamefont {Delugas}, \citenamefont {{DiStasio Jr.}}, \citenamefont
  {Ferretti}, \citenamefont {Floris}, \citenamefont {Fratesi}, \citenamefont
  {Fugallo}, \citenamefont {Gebauer}, \citenamefont {Gerstmann}, \citenamefont
  {Giustino}, \citenamefont {Gorni}, \citenamefont {Jia}, \citenamefont
  {Kawamura}, \citenamefont {Ko}, \citenamefont {Kokalj}, \citenamefont
  {{K\"{u}\c{c}\"{u}kbenli}}, \citenamefont {Lazzeri}, \citenamefont {Marsili},
  \citenamefont {Marzari}, \citenamefont {Mauri}, \citenamefont {Nguyen},
  \citenamefont {Nguyen}, \citenamefont {{Otero-de-la-Roza}}, \citenamefont
  {Paulatto}, \citenamefont {Ponc\'{e}}, \citenamefont {Rocca}, \citenamefont
  {Sabatini}, \citenamefont {Santra}, \citenamefont {Schlipf}, \citenamefont
  {Seitsonen}, \citenamefont {Smogunov}, \citenamefont {Timrov}, \citenamefont
  {Thonhauser}, \citenamefont {Umari}, \citenamefont {Vast}, \citenamefont
  {Wu},\ and\ \citenamefont {Baroni}}]{giannozzi2017advanced}%
  \BibitemOpen
  \bibfield  {author} {\bibinfo {author} {\bibfnamefont {P.}~\bibnamefont
  {Giannozzi}}, \bibinfo {author} {\bibfnamefont {O.}~\bibnamefont
  {Andreussi}}, \bibinfo {author} {\bibfnamefont {T.}~\bibnamefont {Brumme}},
  \bibinfo {author} {\bibfnamefont {O.}~\bibnamefont {Bunau}}, \bibinfo
  {author} {\bibfnamefont {M.}~\bibnamefont {{Buongiorno Nardelli}}}, \bibinfo
  {author} {\bibfnamefont {M.}~\bibnamefont {Calandra}}, \bibinfo {author}
  {\bibfnamefont {R.}~\bibnamefont {Car}}, \bibinfo {author} {\bibfnamefont
  {C.}~\bibnamefont {Cavazzoni}}, \bibinfo {author} {\bibfnamefont
  {D.}~\bibnamefont {Ceresoli}}, \bibinfo {author} {\bibfnamefont
  {M.}~\bibnamefont {Cococcioni}}, \bibinfo {author} {\bibfnamefont
  {N.}~\bibnamefont {Colonna}}, \bibinfo {author} {\bibfnamefont
  {I.}~\bibnamefont {Carnimeo}}, \bibinfo {author} {\bibfnamefont
  {A.}~\bibnamefont {{Dal Corso}}}, \bibinfo {author} {\bibfnamefont
  {S.}~\bibnamefont {{de Gironcoli}}}, \bibinfo {author} {\bibfnamefont
  {P.}~\bibnamefont {Delugas}}, \bibinfo {author} {\bibfnamefont {R.~A.}\
  \bibnamefont {{DiStasio Jr.}}}, \bibinfo {author} {\bibfnamefont
  {A.}~\bibnamefont {Ferretti}}, \bibinfo {author} {\bibfnamefont
  {A.}~\bibnamefont {Floris}}, \bibinfo {author} {\bibfnamefont
  {G.}~\bibnamefont {Fratesi}}, \bibinfo {author} {\bibfnamefont
  {G.}~\bibnamefont {Fugallo}}, \bibinfo {author} {\bibfnamefont
  {R.}~\bibnamefont {Gebauer}}, \bibinfo {author} {\bibfnamefont
  {U.}~\bibnamefont {Gerstmann}}, \bibinfo {author} {\bibfnamefont
  {F.}~\bibnamefont {Giustino}}, \bibinfo {author} {\bibfnamefont
  {T.}~\bibnamefont {Gorni}}, \bibinfo {author} {\bibfnamefont
  {J.}~\bibnamefont {Jia}}, \bibinfo {author} {\bibfnamefont {M.}~\bibnamefont
  {Kawamura}}, \bibinfo {author} {\bibfnamefont {H.-Y.}\ \bibnamefont {Ko}},
  \bibinfo {author} {\bibfnamefont {A.}~\bibnamefont {Kokalj}}, \bibinfo
  {author} {\bibfnamefont {E.}~\bibnamefont {{K\"{u}\c{c}\"{u}kbenli}}},
  \bibinfo {author} {\bibfnamefont {M.}~\bibnamefont {Lazzeri}}, \bibinfo
  {author} {\bibfnamefont {M.}~\bibnamefont {Marsili}}, \bibinfo {author}
  {\bibfnamefont {N.}~\bibnamefont {Marzari}}, \bibinfo {author} {\bibfnamefont
  {F.}~\bibnamefont {Mauri}}, \bibinfo {author} {\bibfnamefont {N.~L.}\
  \bibnamefont {Nguyen}}, \bibinfo {author} {\bibfnamefont {H.-V.}\
  \bibnamefont {Nguyen}}, \bibinfo {author} {\bibfnamefont {A.}~\bibnamefont
  {{Otero-de-la-Roza}}}, \bibinfo {author} {\bibfnamefont {L.}~\bibnamefont
  {Paulatto}}, \bibinfo {author} {\bibfnamefont {S.}~\bibnamefont {Ponc\'{e}}},
  \bibinfo {author} {\bibfnamefont {D.}~\bibnamefont {Rocca}}, \bibinfo
  {author} {\bibfnamefont {R.}~\bibnamefont {Sabatini}}, \bibinfo {author}
  {\bibfnamefont {B.}~\bibnamefont {Santra}}, \bibinfo {author} {\bibfnamefont
  {M.}~\bibnamefont {Schlipf}}, \bibinfo {author} {\bibfnamefont {A.~P.}\
  \bibnamefont {Seitsonen}}, \bibinfo {author} {\bibfnamefont {A.}~\bibnamefont
  {Smogunov}}, \bibinfo {author} {\bibfnamefont {I.}~\bibnamefont {Timrov}},
  \bibinfo {author} {\bibfnamefont {T.}~\bibnamefont {Thonhauser}}, \bibinfo
  {author} {\bibfnamefont {P.}~\bibnamefont {Umari}}, \bibinfo {author}
  {\bibfnamefont {N.}~\bibnamefont {Vast}}, \bibinfo {author} {\bibfnamefont
  {X.}~\bibnamefont {Wu}},\ and\ \bibinfo {author} {\bibfnamefont
  {S.}~\bibnamefont {Baroni}},\ }\bibfield  {title} {\bibinfo {title} {Advanced
  capabilities for materials modelling with {Quantum ESPRESSO}},\ }\href
  {https://doi.org/10.1088/1361-648X/aa8f79} {\bibfield  {journal} {\bibinfo
  {journal} {J. Phys. Condens. Matter}\ }\textbf {\bibinfo {volume} {29}},\
  \bibinfo {pages} {465901} (\bibinfo {year} {2017})}\BibitemShut {NoStop}%
\bibitem [{\citenamefont {Pizzi}\ \emph {et~al.}(2020)\citenamefont {Pizzi},
  \citenamefont {Vitale}, \citenamefont {Arita}, \citenamefont {Bl\"{u}gel},
  \citenamefont {Freimuth}, \citenamefont {G\'{e}ranton}, \citenamefont
  {Gibertini}, \citenamefont {Gresch}, \citenamefont {Johnson}, \citenamefont
  {Koretsune}, \citenamefont {Iba\~{n}ez Azpiroz}, \citenamefont {Lee},
  \citenamefont {Lihm}, \citenamefont {Marchand}, \citenamefont {Marrazzo},
  \citenamefont {Mokrousov}, \citenamefont {Mustafa}, \citenamefont {Nohara},
  \citenamefont {Nomura}, \citenamefont {Paulatto}, \citenamefont {Ponc\'{e}},
  \citenamefont {Ponweiser}, \citenamefont {Qiao}, \citenamefont {Th\"{o}le},
  \citenamefont {Tsirkin}, \citenamefont {Wierzbowska}, \citenamefont
  {Marzari}, \citenamefont {Vanderbilt}, \citenamefont {Souza}, \citenamefont
  {Mostofi},\ and\ \citenamefont {Yates}}]{pizzi2020wannier90}%
  \BibitemOpen
  \bibfield  {author} {\bibinfo {author} {\bibfnamefont {G.}~\bibnamefont
  {Pizzi}}, \bibinfo {author} {\bibfnamefont {V.}~\bibnamefont {Vitale}},
  \bibinfo {author} {\bibfnamefont {R.}~\bibnamefont {Arita}}, \bibinfo
  {author} {\bibfnamefont {S.}~\bibnamefont {Bl\"{u}gel}}, \bibinfo {author}
  {\bibfnamefont {F.}~\bibnamefont {Freimuth}}, \bibinfo {author}
  {\bibfnamefont {G.}~\bibnamefont {G\'{e}ranton}}, \bibinfo {author}
  {\bibfnamefont {M.}~\bibnamefont {Gibertini}}, \bibinfo {author}
  {\bibfnamefont {D.}~\bibnamefont {Gresch}}, \bibinfo {author} {\bibfnamefont
  {C.}~\bibnamefont {Johnson}}, \bibinfo {author} {\bibfnamefont
  {T.}~\bibnamefont {Koretsune}}, \bibinfo {author} {\bibfnamefont
  {J.}~\bibnamefont {Iba\~{n}ez Azpiroz}}, \bibinfo {author} {\bibfnamefont
  {H.}~\bibnamefont {Lee}}, \bibinfo {author} {\bibfnamefont {J.-M.}\
  \bibnamefont {Lihm}}, \bibinfo {author} {\bibfnamefont {D.}~\bibnamefont
  {Marchand}}, \bibinfo {author} {\bibfnamefont {A.}~\bibnamefont {Marrazzo}},
  \bibinfo {author} {\bibfnamefont {Y.}~\bibnamefont {Mokrousov}}, \bibinfo
  {author} {\bibfnamefont {J.~I.}\ \bibnamefont {Mustafa}}, \bibinfo {author}
  {\bibfnamefont {Y.}~\bibnamefont {Nohara}}, \bibinfo {author} {\bibfnamefont
  {Y.}~\bibnamefont {Nomura}}, \bibinfo {author} {\bibfnamefont
  {L.}~\bibnamefont {Paulatto}}, \bibinfo {author} {\bibfnamefont
  {S.}~\bibnamefont {Ponc\'{e}}}, \bibinfo {author} {\bibfnamefont
  {T.}~\bibnamefont {Ponweiser}}, \bibinfo {author} {\bibfnamefont
  {J.}~\bibnamefont {Qiao}}, \bibinfo {author} {\bibfnamefont {F.}~\bibnamefont
  {Th\"{o}le}}, \bibinfo {author} {\bibfnamefont {S.~S.}\ \bibnamefont
  {Tsirkin}}, \bibinfo {author} {\bibfnamefont {M.}~\bibnamefont
  {Wierzbowska}}, \bibinfo {author} {\bibfnamefont {N.}~\bibnamefont
  {Marzari}}, \bibinfo {author} {\bibfnamefont {D.}~\bibnamefont {Vanderbilt}},
  \bibinfo {author} {\bibfnamefont {I.}~\bibnamefont {Souza}}, \bibinfo
  {author} {\bibfnamefont {A.~A.}\ \bibnamefont {Mostofi}},\ and\ \bibinfo
  {author} {\bibfnamefont {J.~R.}\ \bibnamefont {Yates}},\ }\bibfield  {title}
  {\bibinfo {title} {Wannier90 as a community code: {New} features and
  applications},\ }\href {https://doi.org/10.1088/1361-648X/ab51ff} {\bibfield
  {journal} {\bibinfo  {journal} {J. Phys. Condens. Matter}\ }\textbf {\bibinfo
  {volume} {32}},\ \bibinfo {pages} {165902} (\bibinfo {year}
  {2020})}\BibitemShut {NoStop}%
\bibitem [{\citenamefont {Perdew}\ \emph {et~al.}(2008)\citenamefont {Perdew},
  \citenamefont {Ruzsinszky}, \citenamefont {Csonka}, \citenamefont {Vydrov},
  \citenamefont {Scuseria}, \citenamefont {Constantin}, \citenamefont {Zhou},\
  and\ \citenamefont {Burke}}]{perdew2008restoring}%
  \BibitemOpen
  \bibfield  {author} {\bibinfo {author} {\bibfnamefont {J.~P.}\ \bibnamefont
  {Perdew}}, \bibinfo {author} {\bibfnamefont {A.}~\bibnamefont {Ruzsinszky}},
  \bibinfo {author} {\bibfnamefont {G.~I.}\ \bibnamefont {Csonka}}, \bibinfo
  {author} {\bibfnamefont {O.~A.}\ \bibnamefont {Vydrov}}, \bibinfo {author}
  {\bibfnamefont {G.~E.}\ \bibnamefont {Scuseria}}, \bibinfo {author}
  {\bibfnamefont {L.~A.}\ \bibnamefont {Constantin}}, \bibinfo {author}
  {\bibfnamefont {X.}~\bibnamefont {Zhou}},\ and\ \bibinfo {author}
  {\bibfnamefont {K.}~\bibnamefont {Burke}},\ }\bibfield  {title} {\bibinfo
  {title} {Restoring the density-gradient expansion for exchange in solids and
  surfaces},\ }\href {https://doi.org/10.1103/PhysRevLett.100.136406}
  {\bibfield  {journal} {\bibinfo  {journal} {Phys. Rev. Lett.}\ }\textbf
  {\bibinfo {volume} {100}},\ \bibinfo {pages} {136406} (\bibinfo {year}
  {2008})}\BibitemShut {NoStop}%
\bibitem [{\citenamefont {{Van Setten}}\ \emph {et~al.}(2018)\citenamefont
  {{Van Setten}}, \citenamefont {Giantomassi}, \citenamefont {Bousquet},
  \citenamefont {Verstraete}, \citenamefont {Hamann}, \citenamefont {Gonze},\
  and\ \citenamefont {Rignanese}}]{van2018pseudodojo}%
  \BibitemOpen
  \bibfield  {author} {\bibinfo {author} {\bibfnamefont {M.~J.}\ \bibnamefont
  {{Van Setten}}}, \bibinfo {author} {\bibfnamefont {M.}~\bibnamefont
  {Giantomassi}}, \bibinfo {author} {\bibfnamefont {E.}~\bibnamefont
  {Bousquet}}, \bibinfo {author} {\bibfnamefont {M.~J.}\ \bibnamefont
  {Verstraete}}, \bibinfo {author} {\bibfnamefont {D.~R.}\ \bibnamefont
  {Hamann}}, \bibinfo {author} {\bibfnamefont {X.}~\bibnamefont {Gonze}},\ and\
  \bibinfo {author} {\bibfnamefont {G.-M.}\ \bibnamefont {Rignanese}},\
  }\bibfield  {title} {\bibinfo {title} {The {PseudoDojo}: {Training} and
  grading a 85 element optimized norm-conserving pseudopotential table},\
  }\href {https://doi.org/10.1016/j.cpc.2018.01.012} {\bibfield  {journal}
  {\bibinfo  {journal} {Comput. Phys. Commun.}\ }\textbf {\bibinfo {volume}
  {226}},\ \bibinfo {pages} {39} (\bibinfo {year} {2018})}\BibitemShut
  {NoStop}%
\bibitem [{\citenamefont {Nielsen}\ and\ \citenamefont
  {Martin}(1985)}]{nielsen85}%
  \BibitemOpen
  \bibfield  {author} {\bibinfo {author} {\bibfnamefont {O.~H.}\ \bibnamefont
  {Nielsen}}\ and\ \bibinfo {author} {\bibfnamefont {R.~M.}\ \bibnamefont
  {Martin}},\ }\bibfield  {title} {\bibinfo {title} {Quantum-mechanical theory
  of stress and force},\ }\href {https://doi.org/10.1103/PhysRevB.32.3780}
  {\bibfield  {journal} {\bibinfo  {journal} {Phys. Rev. B}\ }\textbf {\bibinfo
  {volume} {32}},\ \bibinfo {pages} {3780} (\bibinfo {year}
  {1985})}\BibitemShut {NoStop}%
\bibitem [{e-i()}]{e-ion-ion}%
  \BibitemOpen
  \href@noop {} {}\bibinfo {note} {Both the electronic ($E_{\text{elec}}$) and
  ion-ion ($E_{\text{ext}}^{\text{I-I}}$) contributions to the total energy are
  required to \emph{quantitatively} reproduce the anomaly. We emphasize
  analysis of $E_{\text{elec}}$ since it can be resolved into orbital-level
  contributions and interpreted chemically. On the other hand,
  $E_{\text{ext}}^{\text{I-I}}$ does not have a localized spatial component but
  is sensitive only to crystal geometry and the effective ionic charge. Care
  must be taken in assessing the geometrical contributions, but by
  pseudopotential choice or construction, the effective ionic charge can be
  fixed across a modest chemical range so that chemistry ($E_{\text{elec}}$)
  can be independently assessed. This aligns with the usual approach in the
  materials physics and chemistry communities, where practioners use,
  \textit{e.g.} COHPs, when a localized picture is needed for chemical
  interpretation.}\BibitemShut {Stop}%
\bibitem [{\citenamefont {Cohen}(1992)}]{cohen92}%
  \BibitemOpen
  \bibfield  {author} {\bibinfo {author} {\bibfnamefont {R.~E.}\ \bibnamefont
  {Cohen}},\ }\bibfield  {title} {\bibinfo {title} {Origin of ferroelectricity
  in perovskite oxides},\ }\href {https://doi.org/10.1038/358136a0} {\bibfield
  {journal} {\bibinfo  {journal} {Nature}\ }\textbf {\bibinfo {volume} {358}},\
  \bibinfo {pages} {136} (\bibinfo {year} {1992})}\BibitemShut {NoStop}%
\bibitem [{\citenamefont {Marzari}\ and\ \citenamefont
  {Vanderbilt}(1998)}]{marzari98}%
  \BibitemOpen
  \bibfield  {author} {\bibinfo {author} {\bibfnamefont {N.}~\bibnamefont
  {Marzari}}\ and\ \bibinfo {author} {\bibfnamefont {D.}~\bibnamefont
  {Vanderbilt}},\ }\bibfield  {title} {\bibinfo {title} {Maximally-localized
  {Wannier} functions in perovskites: Cubic {BaTiO$_3$}},\ }\bibfield
  {journal} {\bibinfo  {journal} {arXiv:cond-mat/9802210}\ }\href
  {https://doi.org/10.1063/1.56269} {10.1063/1.56269} (\bibinfo {year}
  {1998})\BibitemShut {NoStop}%
\bibitem [{\citenamefont {Bhattacharjee}\ and\ \citenamefont
  {Waghmare}(2010)}]{bhattacharjee2010wannier}%
  \BibitemOpen
  \bibfield  {author} {\bibinfo {author} {\bibfnamefont {J.}~\bibnamefont
  {Bhattacharjee}}\ and\ \bibinfo {author} {\bibfnamefont {U.~V.}\ \bibnamefont
  {Waghmare}},\ }\bibfield  {title} {\bibinfo {title} {Wannier orbital overlap
  population ({WOOP}), {Wannier} orbital position population ({WOPP}) and the
  origin of anomalous dynamical charges},\ }\href
  {https://doi.org/10.1039/b918890h} {\bibfield  {journal} {\bibinfo  {journal}
  {Phys. Chem. Chem. Phys.}\ }\textbf {\bibinfo {volume} {12}},\ \bibinfo
  {pages} {1564} (\bibinfo {year} {2010})}\BibitemShut {NoStop}%
\bibitem [{\citenamefont {Stengel}\ \emph {et~al.}(2009)\citenamefont
  {Stengel}, \citenamefont {Spaldin},\ and\ \citenamefont
  {Vanderbilt}}]{stengel09}%
  \BibitemOpen
  \bibfield  {author} {\bibinfo {author} {\bibfnamefont {M.}~\bibnamefont
  {Stengel}}, \bibinfo {author} {\bibfnamefont {N.~A.}\ \bibnamefont
  {Spaldin}},\ and\ \bibinfo {author} {\bibfnamefont {D.}~\bibnamefont
  {Vanderbilt}},\ }\bibfield  {title} {\bibinfo {title} {Electric displacement
  as the fundamental variable in electronic structure calculations},\ }\href
  {https://doi.org/https://doi.org/10.1038/nphys1185} {\bibfield  {journal}
  {\bibinfo  {journal} {Nature Phys.}\ }\textbf {\bibinfo {volume} {5}},\
  \bibinfo {pages} {304} (\bibinfo {year} {2009})}\BibitemShut {NoStop}%
\bibitem [{\citenamefont {Wu}\ and\ \citenamefont {Cohen}(2005)}]{wu05}%
  \BibitemOpen
  \bibfield  {author} {\bibinfo {author} {\bibfnamefont {Z.}~\bibnamefont
  {Wu}}\ and\ \bibinfo {author} {\bibfnamefont {R.~E.}\ \bibnamefont {Cohen}},\
  }\bibfield  {title} {\bibinfo {title} {Pressure-induced anomalous phase
  transitions and colossal enhancement of piezoelectricity in {PbTiO$_3$}},\
  }\href {https://doi.org/10.1103/PhysRevLett.95.037601} {\bibfield  {journal}
  {\bibinfo  {journal} {Phys. Rev. Lett.}\ }\textbf {\bibinfo {volume} {95}},\
  \bibinfo {pages} {037601} (\bibinfo {year} {2005})}\BibitemShut {NoStop}%
\bibitem [{\citenamefont {Ahart}\ \emph {et~al.}(2008)\citenamefont {Ahart},
  \citenamefont {Somayazulu}, \citenamefont {Cohen}, \citenamefont {Ganesh},
  \citenamefont {Dera}, \citenamefont {Mao}, \citenamefont {Hemley},
  \citenamefont {Ren}, \citenamefont {Liermann},\ and\ \citenamefont
  {Wu}}]{ahart08}%
  \BibitemOpen
  \bibfield  {author} {\bibinfo {author} {\bibfnamefont {M.}~\bibnamefont
  {Ahart}}, \bibinfo {author} {\bibfnamefont {M.}~\bibnamefont {Somayazulu}},
  \bibinfo {author} {\bibfnamefont {R.~E.}\ \bibnamefont {Cohen}}, \bibinfo
  {author} {\bibfnamefont {P.}~\bibnamefont {Ganesh}}, \bibinfo {author}
  {\bibfnamefont {P.}~\bibnamefont {Dera}}, \bibinfo {author} {\bibfnamefont
  {H.}~\bibnamefont {Mao}}, \bibinfo {author} {\bibfnamefont {R.~J.}\
  \bibnamefont {Hemley}}, \bibinfo {author} {\bibfnamefont {Y.}~\bibnamefont
  {Ren}}, \bibinfo {author} {\bibfnamefont {P.}~\bibnamefont {Liermann}},\ and\
  \bibinfo {author} {\bibfnamefont {Z.}~\bibnamefont {Wu}},\ }\bibfield
  {title} {\bibinfo {title} {Origin of morphotropic phase boundaries in
  ferroelectrics},\ }\href {https://doi.org/10.1038/nature06459} {\bibfield
  {journal} {\bibinfo  {journal} {Nature}\ }\textbf {\bibinfo {volume} {451}},\
  \bibinfo {pages} {545} (\bibinfo {year} {2008})}\BibitemShut {NoStop}%
\bibitem [{\citenamefont {Zhong}\ \emph {et~al.}(1994)\citenamefont {Zhong},
  \citenamefont {King-Smith},\ and\ \citenamefont {Vanderbilt}}]{zhong94}%
  \BibitemOpen
  \bibfield  {author} {\bibinfo {author} {\bibfnamefont {W.}~\bibnamefont
  {Zhong}}, \bibinfo {author} {\bibfnamefont {R.~D.}\ \bibnamefont
  {King-Smith}},\ and\ \bibinfo {author} {\bibfnamefont {D.}~\bibnamefont
  {Vanderbilt}},\ }\bibfield  {title} {\bibinfo {title} {Giant {LO}-{TO}
  splittings in perovskite ferroelectrics},\ }\href
  {https://doi.org/10.1103/PhysRevLett.72.3618} {\bibfield  {journal} {\bibinfo
   {journal} {Phys. Rev. Lett.}\ }\textbf {\bibinfo {volume} {72}},\ \bibinfo
  {pages} {3618} (\bibinfo {year} {1994})}\BibitemShut {NoStop}%
\end{thebibliography}%

\end{document}